\documentclass[journal,twoside,web]{ieeecolor}
\usepackage{generic}
\usepackage{cite}
\usepackage{amsmath,amssymb,amsfonts}
\usepackage{algorithmic}
\usepackage{mathtools}
\usepackage{textcomp}
\usepackage{multirow}
\usepackage{amsmath}
\usepackage{hyperref}
\usepackage{booktabs}
\usepackage{caption}
\usepackage{soul}
\usepackage{graphicx}
\usepackage{adjustbox}
\usepackage{tablefootnote}
\usepackage[ruled,vlined,linesnumbered]{algorithm2e}
\usepackage{filecontents} 

\definecolor{lightgreen}{RGB}{253, 132, 31}
\definecolor{modification_color}{rgb}{0, 0, 0}
\definecolor{orange}{rgb}{1.0, 0.49, 0.0}
\newcommand\scalemath[2]{\scalebox{#1}{\mbox{\ensuremath{\displaystyle #2}}}}
\usepackage[ruled]{algorithm2e}
\usepackage{amssymb,amsmath,caption}
\usepackage[hang,flushmargin]{footmisc}
\usepackage{lipsum}
\makeatletter
\newcommand{\algorithmfootnote}[2][\footnotesize]{%
  \let\old@algocf@finish\@algocf@finish
  \def\@algocf@finish{\old@algocf@finish
    \leavevmode\rlap{\begin{minipage}{\linewidth}
    #1#2
    \end{minipage}}%
  }%
}

\def\BibTeX{{\rm B\kern-.05em{\sc i\kern-.025em b}\kern-.08em
    T\kern-.1667em\lower.7ex\hbox{E}\kern-.125emX}}
\begin{document}

\thispagestyle{empty}

\begin{table*}[!t]
\vspace{-20cm}
    \centering
    \begin{tabular}{p{16cm}}
    {\LARGE \textbf{IEEE Copyright Notice}} \\
    ~\\
    {\large Copyright © 2023 IEEE} \\
    \\
    {\normalsize Personal use of this material is permitted. Permission from IEEE must be obtained for all other uses, in any current or future media, including reprinting/republishing this material for advertising or promotional purposes, creating new collective works, for resale or redistribution to servers or lists, or
    reuse of any copyrighted component of this work in other works.}
    \end{tabular}
\end{table*}
\newpage

\bstctlcite{IEEEexample:BSTcontrol}
\title{
ORRN: An ODE-based Recursive Registration Network for Deformable Respiratory Motion Estimation with Lung 4DCT Images
}
\author{Xiao Liang, Shan Lin, Fei Liu, Dimitri Schreiber, and Michael Yip, \textit{Senior Member, IEEE}
\thanks{This work was funded by the US Army Telemedicine and Technology Research Center (TATRC)}
\thanks{Xiao Liang is with the Department of Computer Science and Engineering, University of California San Diego, La Jolla, CA 92093 USA. x5liang@ucsd.edu}
\thanks{Shan Lin, Fei Liu, Dimitri Schreiber, and Michael C. Yip are with the Department of Electrical and Computer Engineering, University of California San Diego, La Jolla, CA 92093 USA. {shl102, f4liu, dschreib, yip}@ucsd.edu}}

\maketitle
 
\begin{abstract}
Objective: Deformable Image Registration (DIR) plays a significant role in quantifying deformation in medical data. Recent Deep Learning methods have shown promising accuracy and speedup for registering a pair of medical images. However, in 4D (3D + time) medical data, organ motion, such as respiratory motion and heart beating, can not be effectively modeled by pair-wise methods as they were optimized for image pairs but did not consider the organ motion patterns necessary when considering 4D data.
Methods: 
This paper presents ORRN, an Ordinary Differential Equations (ODE)-based recursive image registration network. Our network learns to estimate time-varying voxel velocities for an ODE that models deformation in 4D image data. It adopts a recursive registration strategy to progressively estimate a deformation field through ODE integration of voxel velocities.
Results:
We evaluate the proposed method on two publicly available lung 4DCT datasets, DIRLab and CREATIS, for two tasks: 1) registering all images to the extreme inhale image for 3D+t deformation tracking and 2) registering extreme exhale to inhale phase images. Our method outperforms other learning-based methods in both tasks, producing the smallest Target Registration Error of 1.24mm and 1.26mm, respectively. Additionally, it produces less than 0.001\% unrealistic image folding, and the computation speed is less than 1 second for each CT volume.
Conclusion: 
ORRN demonstrates promising registration accuracy, deformation plausibility, and computation efficiency on group-wise and pair-wise registration tasks.
Significance: It has significant implications in enabling fast and accurate respiratory motion estimation for treatment planning in radiation therapy or robot motion planning in thoracic needle insertion.
\end{abstract}


\section{Introduction}
\label{sec:introduction}
 
\IEEEPARstart{A} primary goal in radiation therapy is to maximize the radiation dosage provided to the target tissue while minimizing the radiation exposure of other nearby organs at risk. However, respiratory motion can cause displacement of both the targeted lesion and the surrounding anatomy in areas such as the lung and liver. It needs to be accurately accounted for in treatment planning and dose delivery during radiation therapy in many critical organs \cite{doi:10.3109/0284186X.2013.813638}, \cite{yang2008}. As a solution, four-dimensional computed tomography (4DCT) can be used to derive valuable motion information about the target tumor and other normal tissue. Previous studies have utilized conventional DIR techniques with 4DCT to deliver precise radiation to the target tumor and avoid unwanted doses to nearby organs at risk\cite{Flampouri_2006}, \cite{10.1371/journal.pone.0271064}. Although conventional DIR methods provide accurate deformation estimation, they have a long computation time that can take hours per scan due to iterative optimization processes. This prolonged computation time makes them unsuitable for clinical applications requiring online computation. In contrast, current Deep Learning-based DIR methods \cite{Hansen2021}, \cite{HERING2021102139} had shown comparable registration accuracy to conventional methods while requiring much less computation time.  

\begin{figure}[!t]
\captionsetup{margin=0.1cm}
\centering
\includegraphics[width=0.45\textwidth]{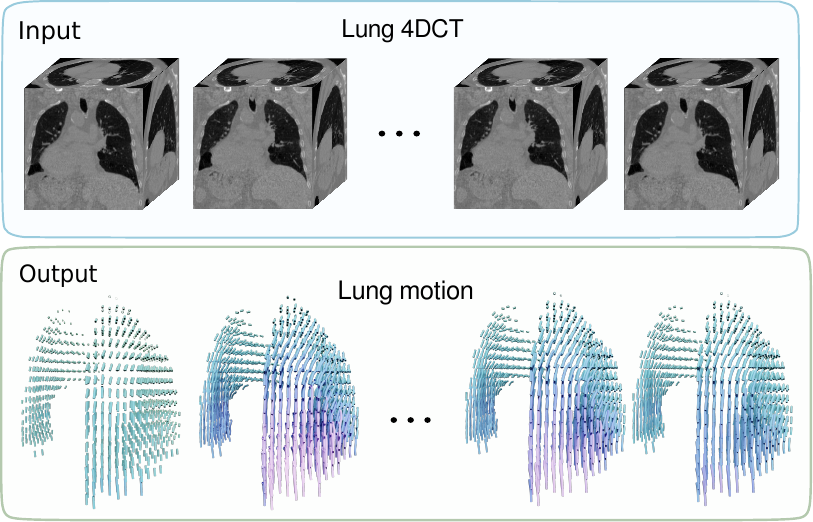}

\caption{A visualization of the proposed DIR method applied to a respiratory motion estimation task. It takes a 4DCT lung scan and estimates dense displacement fields through ODE integration. Calculated displacement fields are applied back to the initial CT image to achieve lung motion estimation.}
\label{cover}
\end{figure}
Until now, most learning-based methods focusing on the thoracic region have only addressed DIR between a pair of scans and have ignored the temporal continuity in 4DCT. Although a few group-wise registration methods \cite{Zhang_2021,Lu_2021} have attempted to register all 4DCT scans to a common template image, they were limited by slow test-time optimization or insufficient registration accuracy. In addition, previous learning-based methods predicted direct spatial transformation between discrete phase images in 4DCT but neglected intermediate changes.

To solve the problems above, we propose a new unsupervised learning-based DIR approach that learns to estimate time-varying voxel velocities in 4DCT scans for an ODE that models the lung deformation throughout a breathing cycle. We adapt a recursive decomposition strategy where an estimation of a large deformation is broken down into multiple velocity estimations and ODE integration steps. With our ODE-based formulation and the accompanying recursive strategy, we construct an ODE-based Recursive Image Registration Network (ORRN) that has advantages at modeling dense lung motion and tracking large deformation. The proposed work is trained in a fully unsupervised manner, bypassing challenges associated with building labelled datasets and further generalizing the method. We evaluate the effectiveness of the proposed method in both group-wise and pair-wise registration on two publicly available 4DCT datasets. In summary, our main contributions are
\begin{itemize}
    \item We propose a novel ODE-based Recursive Registration Network (ORRN) that progressively estimate deformation through an ODE integration of predicted voxel velocities at any given time in 4DCT. 

    \item The proposed method reaches state-of-the-art (SOTA) accuracy in a group-wise registration scenario. It generates voxel trajectory for respiratory motion estimation.
    
    \item We show that by combining the multi-scale and recursive strategies, our method also outperforms previous unsupervised learning methods for pair-wise registration of lung 4DCT. Particularly, it shows promising ability at dealing with large deformation accurately.

    \item We conduct extensive experiments and ablation studies to validate effectiveness and efficiency of the proposed ORRN. 
    
\end{itemize}
 Related works are discussed in the next section (Section \ref{method}). \textcolor{modification_color}{In Section \ref{method}}, we describe our problem formulation, neural network architecture, data flow, and training loss functions. \textcolor{modification_color}{In Section \ref{result_and_exp}}, we present experiments and results, including CT data used in this study, related implementation details, results of registration accuracy and deformation plausibility, and ablation studies. In Section \ref{sec:discussion}, we discuss the proposed approach's advantages and limitations. Lastly, we conclude our work and future directions.

\section{Related work}
Recent comprehensive reviews of DIR algorithms can be found in \cite{Fu2020a}, \cite{Sotiras2013}, and \cite{VIERGEVER2016140}. In the following sections, we provide overviews of conventional methods, deep learning-based methods, deformation decomposition strategies, and the Neural Ordinary Differential Equations (Neural ODEs) framework that is used in this work.

\subsubsection{Conventional Methods}
Conventional methods treat DIR as an optimization problem. They iteratively optimize an energy function that often consists of an image-matching term, such as sum-of-squared differences and local correlation coefficient, measuring alignment between a deformed image and a fixed image, and a regularization term that encourages a smooth, realistic deformation. Popular methods include free-form deformations with cubic B-splines \cite{ffd2}, elastic body model\cite{elastic2}, diffusion model \cite{demon} and physical simulation-based model \cite{Ladjal2021}, \cite{Fei_2021_ICRA}. Besides registering \textcolor{modification_color}{pairs of images}, conventional group-wise algorithms \cite{popi}, \cite{Metz2011} registered all images to a common template utilizing temporal constraints such as cyclic motion prior. \cite{8360027} used a sparse-to-dense approach relying on a pre-computed patient-specific model to speed up computation. Xue \emph{et al.}\cite{9220764} modeled lung motion by applying a fast Kalman filter on two previous conventional DIR methods that were run on GPUs. \textcolor{modification_color}{In general, conventional optimization-based DIR methods have high registration accuracy. Their main drawbacks are long computation time and the potential demand for parameter adjustments in order to achieve good accuracy \cite{Song2017ARO}.}

\subsubsection{Deep Learning Methods}
Many Deep Learning-based algorithms for DIR have emerged in recent years. In general, they are fast but still suffer from inaccurate registration results due to the requirement of a large training dataset. Supervised learning methods such as \cite{8902170}, \cite{gdlfire}, and \cite{modelbasedartificial} trained neural networks with deformation ground truth generated using artificial motion or conventional DIR methods' output. They were not accurate because of the gap between synthetic deformation and real deformation \cite{Fu2020}. To circumvent the deficiency in deformation ground truth, unsupervised methods used a spatial transformer \cite{spatialtransformer} for image warping and formulated learning objectives with image similarity metrics similar to those used in conventional methods. Commonly-used designs include UNet-based architectures \cite{Balakrishnan2018}, \cite{mlvirnet}, key points-based methods \cite{cueaware}, \cite{Hansen2021}, adversarial learning \cite{Fu2020}, \cite{Yang_2021}, recurrent neural network networks \cite{Lu_2021}, \cite{recurrentregistration}, and various decomposition methods \cite{recursive2}, \cite{Mok2020}, \cite{Zhao2019} (discussed in the next section). Currently, the best pair-wise registration accuracy on Lung CT data was achieved by an unsupervised method \cite{Hansen2021} that used a Graph-based network for key points displacement prediction and a weakly-supervised method \cite{HERING2021102139} that used key points and lung mask-based loss functions in addition to image-based metrics during training. Group-wise registration for 4D medical data is relatively less explored by learning-based methods. One-shot and few-shot learning methods \cite{fechter2019, Zhang_2021, Chi_2022} performed group-wise registration by optimizing their network parameters for every new data. Lu \emph{et al.} \cite{Lu_2021} proposed a convolutional recurrent neural network to perform group-wise registration for 4D lung CT. Their method could generalize to unseen data, but its accuracy still needed to be improved.

\subsubsection{Decomposition in Learning-based DIR}
In this work, we define decomposition in DIR as breaking down a large deformation into estimation of several smaller deformations. In that sense, previous works \cite{Mok2020}, \cite{HERING2021102139}, \cite{Jiang2020}, \cite{He2021} with a multi-scale architecture, that progressively refine a predicted deformation at each level, belonged to this category. Such methods were efficient because the low-resolution components were good at capturing large deformation due to sufficient receptive fields, and high-resolution components only needed to refine details. However, errors made at each level were propagated to the next level as well which made such design unstable. Another decomposition strategy is the recursive strategy, in which the same network component (with or without weight sharing), is ran multiple times recursively. At each time, it predicts a small deformation based on a previous progressively accumulated deformation. These include methods like \cite{Zhao2019}, and \cite{recursive1}. Note that they differed from conventional methods because the later iteratively optimize deformation parameters whereas learning-based recursive methods progressively predicted deformation by a certain number of neural networks' forward passes. Most recently, Hu \emph{et al.}\cite{recursive2} unified the multi-scale and recursive approaches by applying recursion on multiple resolution levels and stages. Similarly, our work takes advantages of both decomposition strategies to accomplish efficient and accurate DIR, but it formulated recursion as an ODE integration of predicted time-varying voxel velocity, which distinguishes it from previous works.

\subsubsection{Neural ODEs}
Neural ODEs was first introduced by Chen \emph{et al.}\cite{Chen2018}. Aiming to simulate an infinite-depth neural network, the framework used a single neural network to parameterize the derivative of the hidden state. It has shown promising results in areas such as normalizing flow \cite{Chen2018} and learning particle dynamics \cite{Niemeyer2019ICCV}, etc. Recently, Xu \emph{et al.}\cite{Xu2021} introduced Neural ODEs to the field of DIR, using multi-scale Neural ODEs to learn a continuous optimizer. Similar to a conventional approach, Wu \emph{et al.}\cite{wu2021} used a convolutional network with Neural ODEs to optimize the trajectory of voxels during inference. In our work, \textcolor{modification_color}{Neural ODEs are used as integration frameworks.} Because we let a neural network estimate voxel velocities and integrate through them over time, our method fits into the Neural ODEs framework. 

\begin{table}[h!]
\setlength\tabcolsep{1.5em}
\centering
\caption{\textcolor{modification_color}{General Notations and definitions}}\label{notations}
\begin{adjustbox}{width=0.48\textwidth}
\begin{tabular}{lc}
\toprule

 Notation  & Definition \\ \midrule
$\mathbf{I}$    & 3D CT scan \\ 
$\mathcal{F}$   & Image feature \\ 
$\Phi$  & Dense displacement field  \\
Subscript $t$   & a variable at time point $t$ \\ 
Subscript $m, f, w$   & Moving (m) or fixed (f) or warped (w) \\
$\prescript{i}{}{F}$  & $i$-th element of a list $F$ \\
$\lceil d \rceil$  & Ceiling for a decimal value $d$ \\
$\mathcal{F} \circ \Phi$  &  Spatially warp $\mathcal{F}$ with $\Phi$\\
$\mathbf{p} \oplus \Phi$  & Apply displacement $\Phi$ to $\mathbf{p} \in \mathbb{R}^3$ \\
\bottomrule
\end{tabular}
\end{adjustbox}
\end{table}

\section{Methods}\label{method}
 
\subsection{Problem Formulation}\label{sec:problem_formulation}
\textcolor{modification_color}{Some general notations and definitions are provided in Table \ref{notations}}. Let $\mathbf{I}^T = \{ \mathbf{I}_t |t = 0,\cdots,T-1\}$ \textcolor{modification_color}{denote} a sequence of 3D CT scans with length $T$, where each $\mathbf{I}_t: H\times W \times D \rightarrow \mathbb{R}$ maps the 3D voxelgrid into gray-scale image intensity.
In this work, we model deformation as voxel displacement from a fixed image $\mathbf{I}_f$ with a dense displacement field $\Phi \in \mathbb{R}^3$. As an ODE is an equation that is defined for a function of one independent variable and its derivatives, the evolution of $\Phi$ can be formulated as an ODE:
\begin{equation}
\label{eqn:dphidt}
\begin{split}
    \frac{d}{dt}\Phi_t = \mathbf{v}(\Phi_t, t), \Phi_0 = \mathtt{zeros}
\end{split}
\end{equation}
where $\mathbf{v}(\cdot)$ is a function that predicts voxel velocity $\dot{\Phi_t}$, and $\Phi_0$ is initialized with zeros. In this work, we focus on designing $\mathbf{v}(\cdot)$ as a deep learning module to perform accurate registration. We regard the extreme 
inhale (EI) phase image as the fixed image $\mathbf{I}_f$ and focus on two separate tasks:
\subsubsection{Group-wise Registration}\label{pw4dformulation}
In this task, we aim to model the respiratory motion through group-wise registration of 4D CT scans. The network learns to predict a set of dense displacement fields ${\Phi^T=\{\Phi_t}|t=0,\cdots,T-1\}$ representing voxel trajectories over time. A displacement field at any time point $t$ can be obtained by integration:
\begin{equation}
\begin{split}\label{eqn:integral_group}
        \Phi_{T}& =  \Phi_0 + \int_{0}^{T}\mathbf{v}(\Phi_t, t)dt \\
\end{split}
\end{equation}
such that moving CT images $\mathbf{I}_t$ can be warped using $\Phi_t$ to align with the fixed image $\mathbf{I}_f$. \textcolor{modification_color}{In practice, we use Euler discretization to approximate the ODE:
\begin{equation*}
\begin{split}\label{eqn:euler_discretization}
        \Phi_{t+h}& \approx  \Phi_t + \mathbf{v}(\Phi_t, t) \cdot h \\
\end{split}
\end{equation*}
where $h$ is a step size. Equation \ref{eqn:integral_group} can then be approximated by discretizing the continuous time to $K$ intervals:
\begin{equation*}
\begin{split}\label{eqn:euler_discretization}
        \Phi_{T}& \approx  \Phi_0 + \sum_{k=0}^{K-1} \mathbf{v}(\Phi_{k \times h}, k \times h) \cdot h \\
\end{split}
\end{equation*}}

\subsubsection{Pair-wise Registration}
In addition to performing group-wise registration, our method also serves as a general DIR framework that can be utilized for pair-wise registration. We operate under the assumption that only the images representing the extreme inhale (EI) and extreme exhale (EE) phases are accessible. The proposed network learns to estimate a displacement field $\Phi_1$:
\begin{equation}
\label{eqn:integral_pair}
\begin{split}
    \Phi_1 & =  \Phi_0 + \int_{0}^{1}\mathbf{v}(\Phi_t, t)dt \\
    & \textcolor{modification_color}{\approx  \Phi_0 + \sum_{k=0}^{K-1} \mathbf{v}(\Phi_{k \times h}, k \times h) \cdot h }
\end{split}
\end{equation}
such that $\Phi_1$ deforms the EE phase image $\mathbf{I}_m$ to be aligned with the EI phase image $\mathbf{I}_f$. Because we don't consider time in a pair-wise scenario, we always integrate an ODE from 0 to 1.

 \begin{algorithm}[h!]
 \small
    \caption{Pseudo algorithm for ORRN}
    \label{alg:orrn}
    \SetKwInOut{Input}{Input}
    \SetKwInOut{Output}{Output}
    \Input{4D CT scans $\mathbf{I}^T=\{\mathbf{I}_t|t=0,\cdots,T-1\}$, or 3D CT pairs $\mathbf{I}^T= \{\mathbf{I}_f, \mathbf{I}_m\}$, where $T$ is the length of $\mathbf{I}^T$.}
    $h, F \gets \text{zeros}, [\ ]$ \\
    \tcp{\footnotesize \textcolor{lightgreen}{Encode image and temporal features}}
    \For{$\mathbf{I}$ in $\mathbf{I}^T$} {
        {$\mathcal{F}$ $\gets$ {$\textbf{FeaturePyramid}(\mathbf{I})$}}\\
        {$h$ $\gets$ {$\textbf{ConvGRU}(\mathcal{F}, h)$}}\\
        {$F \gets [F, \mathcal{F}]$}\\
    }
    $\Phi^T = [\mathtt{zeros}]$ \\
    \For{$i \gets 1$ to $T$} {
        {$\mathcal{F}_f, \mathcal{F}_m, \mathcal{C} \gets \prescript{0}{}{F}, \prescript{i}{}{F}$, $h$ }\\
        {$input \gets [\mathcal{F}_f, \mathcal{F}_m, \mathcal{C}, \prescript{(i-1)}{}{\Phi^T}]$}\\
        \tcp{\footnotesize \textcolor{lightgreen}{integrate from $t-1$ to $t$}}
        { $\Phi^T \gets [\Phi^T, \textbf{ODEInt}(\text{VelocityEstimation}, input, [i-1, i])]$}
    }
     \Return{$\Phi^T$}
\end{algorithm}

\begin{figure*}[t!]
\centering
\includegraphics[width=0.95\linewidth]{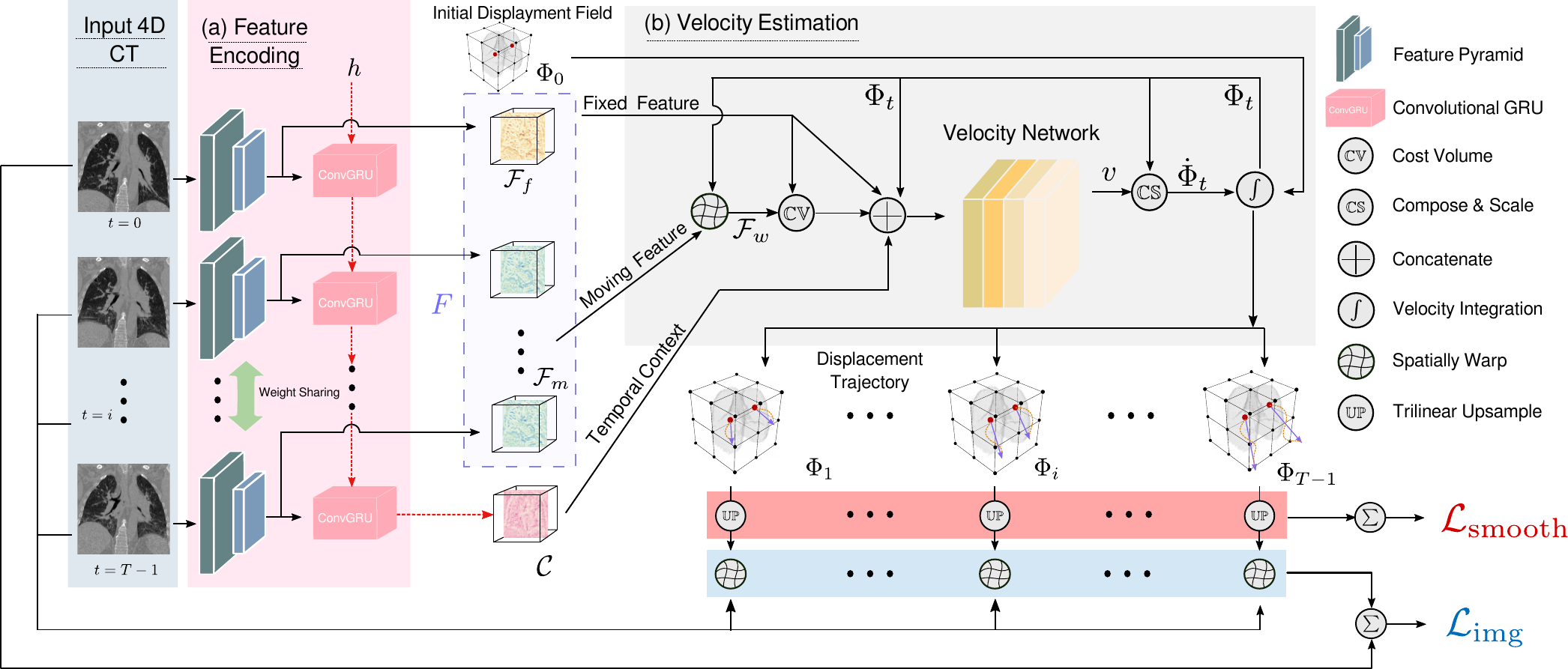}
\caption{Overview of the proposed ODE-based Recursive Registration Network. 4DCT scans are first encoded by a feature pyramid followed by a ConvGRU unit to obtain individual feature maps $\mathcal{F}_f, \mathcal{F}_m$ and the temporal context feature $\mathcal{C}$. Starting with the zero displacement, our method predicts voxel velocities with the velocity estimation module. Every step, a selected moving feature $\mathcal{F}_m$ is warped by the current displacement $\Phi_t$ to compute a Cost Volume with a fixed feature $\mathcal{F}_f$.\ \ $\mathcal{F}_f, \Phi_t, \mathcal{C}, \text{and the Cost Volume}$ are concatenated to construct input for the velocity network. The network outputs a vector field $v$, which is then converted to voxel velocity $\dot{\Phi}_t$ through a \textit{Compose\&Scale} operation. Finally, ODE integration produces a set of dense displacements \textcolor{modification_color}{$\{\Phi_0,\Phi_1...\Phi_{T-1}\}$} which is then used for loss calculation. Note that when performing pair-wise registration, the input to the network will be only a pair of images and output will be a set of displacement fields $\Phi_t$, where $t \in [0,1]$.}\label{workflow_123}
\end{figure*}

\subsection{Learning-based Framework}\label{net_architecture}

\subsubsection{Overall Network Design}
Figure \ref{workflow_123} and Algorithm \ref{alg:orrn} show the workflow of the proposed framework. It consists of (a) the feature encoding stage and (b) the ODE integration stage. During feature encoding, a feature encoding network is used for extracting individual image features and the temporal context of 4DCT. Conditioning on the output from feature encoding, a velocity estimation module predicts voxel velocities at every integration time point during the integration stage. Specifically, in group-wise registration task, a CT sequence is encoded by the feature encoder, and during integration, we model voxel displacement starting from the EI image throughout the whole CT sequence. In the pair-wise registration scenario, only a pair of images are encoded and voxel displacements are modeled from the fixed image to a moving image. For more details on the neural network architecture, see Appendix \ref{netdetails}.

\subsubsection{Feature Encoding}
 The feature encoding pipeline is visualized in part (a), \textcolor{modification_color}{Figure \ref{workflow_123}}. A feature pyramid network is used for extracting individual image features for each CT image using shared weight. It has two levels. Each level encodes input to higher feature dimensions while reducing each spatial resolution by a factor of two, generating feature maps at half and quarter of input resolution. Its architecture is adapted from the Recursive Refinement Network (RRN)\cite{He2021} by taking only the first two levels of its Siamese feature encoder. After extracting individual image features, a Convolutional Gated Recurrent Unit (ConvGRU)\cite{cgru} encodes temporal information features extracted by the feature pyramid. Specifically, outputs of the last feature pyramid level and a previous hidden state vector (initialized with zeros) are fed to the ConvGRU that will produce a new hidden state. In summary, the feature encoding stage will produce a set of image feature maps $\mathcal{F}_s$, whose each entry is a multi-resolution feature map for a CT image, and a temporal context $\mathcal{C}$ that is the final hidden state of the GRU.

\subsubsection{Velocity Estimation}
We tackle the problem of large deformation by guiding the velocity prediction module with the next adjacent frame of the current time point. To estimate time-varying voxel velocity, we propose a velocity estimation module that is guided by the next adjacent feature from the current time point (Figure \ref{workflow_123}, part (b)). It predicts a change $\dot{\Phi}_t$ to the previous voxel displacement field $\Phi_t$. In this section, we discuss feature maps and velocity prediction on a single-resolution (a quarter of input resolution) for simplicity. A more general and accurate network design that uses multi-resolution architecture will be discussed in the next section.

At every integration time point $t$, we define the current fixed and moving image features as $\mathcal{F}_f=\prescript{0}{}F, \mathcal{F}_m=\prescript{\lceil t \rceil}{}F$. The proposed module takes $\mathcal{F}_f, \mathcal{F}_m$, $\mathcal{C}$, and $\Phi_t$ as input.
In each estimation step, $\Phi_t$ will be used to warp $\mathcal{F}_m$ with a spatial transformer\cite{spatialtransformer} to get a warped feature $\mathcal{F}_w$. A local cost volume $\scalemath{0.8}{\mathbb{CV}}$ \cite{He2021}, \cite{uflow} is computed between $\mathcal{F}_w$ and $\mathcal{F}_f$ to provide additional correspondence information (for details about cost volume, please refer to Appendix \ref{localcostvolume}). A velocity network \textbf{VN} directly predicts a vector field $v$ that approximates a deformation from warped features $\mathcal{F}_w$ to fixed features $\mathcal{F}_f$. The vector field is then converted to the voxel velocity $\dot \Phi_t$ by a \textit{Compose\&Scale} ($\scalemath{0.8}{\mathbb{CS}}$) operation. In this operation, the vector field $v$ is converted by composing on and subtracting from $\Phi_t$. Then, the velocity's magnitude is scaled by dividing the remaining time from the current integration time points to time points of the current moving image, which encourages predicted velocities to have constant magnitudes (for details, please refer to Appendix \ref{composeandscale}). Lastly, the voxel velocity $\dot \Phi_t$ is used for ODE integration similar to equation \ref{eqn:dphidt}, obtaining a trajectory of $\Phi_t$.

\subsubsection{Multi-resolution Architecture}

Our experiments show that our ORRN with a single-resolution velocity estimator already outperforms comparison methods. We demonstrate that it can be extended to using a multi-resolution architecture that recursively predicts and refines the vector field $v$ at different resolutions to get better accuracy. Let $L=\{l_i|i=1,...,n\}$ be a set of fractional numbers in an ascending order, where each of its elements represents the ratio of a feature map's spatial dimension to the original CT dimension. We now indicate that a variable is in a certain spatial resolution with superscript $l_i$ (e.g. $\mathcal{F}_w^{l_i}$). A velocity network component that solely operates at a certain resolution is labeled similarly (e.g. $\textbf{VN}^{l_i}$). The general flow of a multi-resolution velocity network can be summarized as

\begin{equation}
\begin{aligned}
    &\ \ \ \ v^{l_0} =\text{zeros},\ \phi^{l_n} = \Phi_{t}^{l_n},\ \mathcal{C}^{l_0}=\mathcal{C}\\
   \text{repeat} & \begin{dcases}
            \mathcal{C}^{l_i}, v^{l_i}, \phi^{l_i} &= \textit{Resample}(\mathcal{C}^{l_{i-1}}, v^{l_{i-1}}, \phi^{l_n})\\
    \hfil  \phi' &= \phi^{l_i} \circ  v^{l_i} \\
    \hfil  \mathcal{F}_w^{l_i} &= \mathcal{F}_m^{l_i} \circ  \phi'\\
     \hfil  \text{CV}^{l_i} &= \mathbb{CV}(\mathcal{F}_f^{l_i}, \mathcal{F}_w^{l_i}) \\
    \hfil  v^{l_i}, C^{l_i} & = \textbf{VN}^{l_i}(\text{CV}^{l_i}, \phi', \mathcal{F}_f^{l_i}, \mathcal{C}^{l_i})\\
    \end{dcases}\\
&\ \ \ \ \dot{\Phi}^{l_n}_{t} ={\mathbb{CS}}( v^{l_n}, \Phi_{t}^{l_n}), 
\end{aligned}
\end{equation}where $\textit{Resample}(\cdot)$ indicates a interpolation operation. Lines that are included by the curly bracket will be repeated n times for refining the predicted vector field at multiple resolution levels. Note that because the context feature $\mathcal{C}$ encoded by the ConvGRU unit is initially only available at the lowest resolution, it is later replaced with the outputs of the current network component's second last layer. In our experiment, we use resolution levels $\{l_1:\frac{1}{4}\}$ for the group-wise registration task. For pair-wise registration, we explore both $\{l_1:\frac{1}{4}\}$ and $\{l_1:\frac{1}{4},\ l_2:\frac{1}{2}\}$ setting. We use a small number of resolution levels mainly due to computational limitations. A possible future direction is to adopt lightweight network architectures to reduce computational burdens, which may make further applying the multi-resolution structure possible.

\subsection{Loss functions}
\label{lossf}
The overall loss function $\mathcal{L}_{total}$ used in this study consists of two terms, a image dissimilarity term $\mathcal{L}_{img}$ that encourages alignment between two images and a smoothness term $\mathcal{L}_{smooth}$ for regularization:

\begin{equation}
\small
\begin{split}
   & \textstyle \mathcal{L}_{total} = \\
    & \begin{dcases}
      \frac{1}{T}\sum_{t=1}^{T}\Bigl( \mathcal{L}_{img}(\mathbf{I}_f, \mathbf{I}_t\circ  \Phi_t ) +  \mathcal{L}_{smooth}(\Phi_t) \Bigr),& \text{Group-wise} \\
            \mathcal{L}_{img}(\mathbf{I}_f, \mathbf{I}_m \circ \Phi_1 ) +  \mathcal{L}_{smooth}(\Phi_1),& \text{Pair-wise} \\
    \end{dcases}
\end{split}
\end{equation}


\subsubsection{Image Loss}
$\mathcal{L}_{img}$ is calculated as the negative normalized local cross-correlation \cite{Balakrishnan2018},\cite{AVANTS200826}. Let $\mathbf{I}, \mathbf{J}$ be two input gray-scale image volumes of the same shape, then let $\mathbf{\hat I}, \mathbf{\hat J}$ be their local intensity images where the local mean at position $\mathbf{p}$, $\mathbf{{\hat I}_p}$ or $\mathbf{{\hat J}_p}$, is computed over a cubic window $W$ with a width of $n$ centered at position $\mathbf{p}$ and $\mathbf{p_i}$ is an element within the window $W$. In this work, $n=9$. The image loss can be written as:
\begin{equation}
\small
  \label{eqn:cc}
   \mathcal{L}_{img}(\mathbf{I}, \mathbf{J}) = -\sum_{\mathbf{p} \in \Omega} \frac{ \left(\sum_{\mathbf{p_i}}(\mathbf{I_{p_i}}-\mathbf{{\hat I}_p})(\mathbf{J_{p_i}} - \mathbf{{\hat J}_p}) \right)^2}{\sum_{\mathbf{p_i}}(\mathbf{I_{p_i}}-\mathbf{{\hat I}_p})^2\cdot\sum_{\mathbf{p_i}}(\mathbf{J_{p_i}} - \mathbf{{\hat J}_p})^2}
\end{equation}
\subsubsection{Smoothness Loss}
For encouraging local smoothness in dense displacement fields, we regularize its spatial gradient. $\mathcal{L}_{smooth}$ is defined as the mean spatial gradient of a deformation field\cite{Balakrishnan2018}:
\begin{equation}
  \label{eqn:smooth}
  \mathcal{L}_{smooth}(\Phi) =\sum_{\mathbf{p}\in \Omega} \|\nabla\Phi(\mathbf{p})  \|_2^2,
\end{equation}
where $\nabla\Phi(\mathbf{p})$ denotes the spatial gradient at position $\mathbf{p}$.

\begin{table*}[!htbp]
\setlength\tabcolsep{2.4em}
\centering
\captionsetup{justification=centering,margin=0.5cm}
\caption{Target Registration Error of comparison methods for \textcolor{modification_color}{Task A}. Mean TRE is reported for 750 landmarks in DIR-Lab 4DCT labeled on all exhalation phase images(0 to 50\%) and for 300 landmarks in CREATIS dataset labeled on all 10 phase images. Statistical significance level are defined as: $^{*}$ for $p < 0.05$,  $^{**}$ for $p < 0.01$, $^{***}$ for $p < 0.001$}
\begin{adjustbox}{width=0.9\textwidth}
\begin{tabular}{c|c|ccc|ccccc}
\toprule
 \multicolumn{1}{c}{}& \multicolumn{1}{c}{}  &\multicolumn{3}{c}{Non-learning method} & \multicolumn{5}{c}{Unsupervised learning method}     \\ \midrule
          & W/o reg &Me11   &LRME & FEM & LCRN & 
    RRN    & LapIRN     & GRN & ORRN$^1$ \\ \hline
4DCT 01   & $2.18$    & $0.95$ & $0.58$ & $1.80$    & $1.07$ &  $0.99$  & \boldmath$0.95$ & $0.96$  &$0.96$ \\
4DCT 02   & $3.78$    & $1.00$ & $0.63$ & -          & $1.02$  &  $1.02$ & $0.98$ & \boldmath$0.95$  & $0.99$ \\
4DCT 03   & $5.05$    & $1.14$ & $0.83$ &  -        & $1.17$  &  $1.14$ & $1.14$          & $1.13$  & \boldmath$1.11$ \\
4DCT 04   & $6.69$    & $1.40$ & $1.11$ & $2.00$     & $1.49$  &  $1.37$ & $1.33$          & $1.33$  & \boldmath$1.31$  \\
4DCT 05   & $5.22$    & $1.50$ & $1.02$ &   -        & $1.40$  &  $1.41$ & $1.38$          & $1.35$  & \boldmath$1.34$ \\
4DCT 06   & $7.42$    &   -    & $1.14$ & $2.00$    & $1.58$  &  $1.61$ & $1.61$          & $1.62$  & \boldmath$1.52$  \\
4DCT 07   & $6.66$    &   -    & $0.93$ &   -        & $1.43$ &  $1.51$ & $1.38$          & $1.33$  & \boldmath$1.29$ \\
4DCT 08   & $9.83$    &   -    & $1.10$ &   -      & $2.12$  &  $2.08$ & $1.98$          & $1.56$  & \boldmath$1.52$ \\
4DCT 09   & $5.03$    &   -    & $0.90$ & $1.90$   & $1.28$   &  $1.26$ & \boldmath$1.16$& $1.22$ & $1.17$ \\
4DCT 10  & $5.42$     &   -    & $0.93$ & $1.80$    & $1.39$   &  $1.33$ & $1.37$          & $1.38$ & \boldmath$1.23$ \\ \midrule
Avg   & $5.73$        & $1.20$ & $0.92$  & $1.9$     & $1.39$ &  $1.37$ & $1.33$ & $1.28$ & \boldmath$1.24$ \\
std   & $5.65$          &  -    &   -       &   -   & $0.98$ & $1.34$  & $1.40$ & $1.08$ & $1.05$ \\
Sig level   &   ***       &  -  & - & -   & - & *** & *  & ** &       \\
\bottomrule \toprule
CREATIS 01  & $3.58$ & - & $1.00$   & - & - &$1.12$ & $1.09$   & \boldmath$1.06$ & $1.10$   \\
CREATIS 02  & $9.98$ & - & $1.36$   & - & - &$1.69$ & $1.68$   & $1.57$ & \boldmath$1.49$  \\
CREATIS 03  & $5.28$ & - & $0.95$   & - & - &$1.15$ & \boldmath$1.07$   & $1.14$ & $1.09$  \\ \midrule

Avg   & $6.28$           &    -            & -       & -     &-  & $1.32$ & $1.28$ & $1.26$ & \boldmath$1.23$ \\
std   & $5.44$           &    -             &  -      & -     &-  & $0.96$ & $1.01$ & $0.88$ & $0.81$ \\
Sig level   &  ***        &  -  & - &  -  & - & *** & /  & /  &       \\
\bottomrule
\end{tabular}
\end{adjustbox}

\label{4dtable}

\end{table*}
\begin{figure*}[h!]
\captionsetup{margin=0.1cm}
\centering
\includegraphics[width=0.87\textwidth]{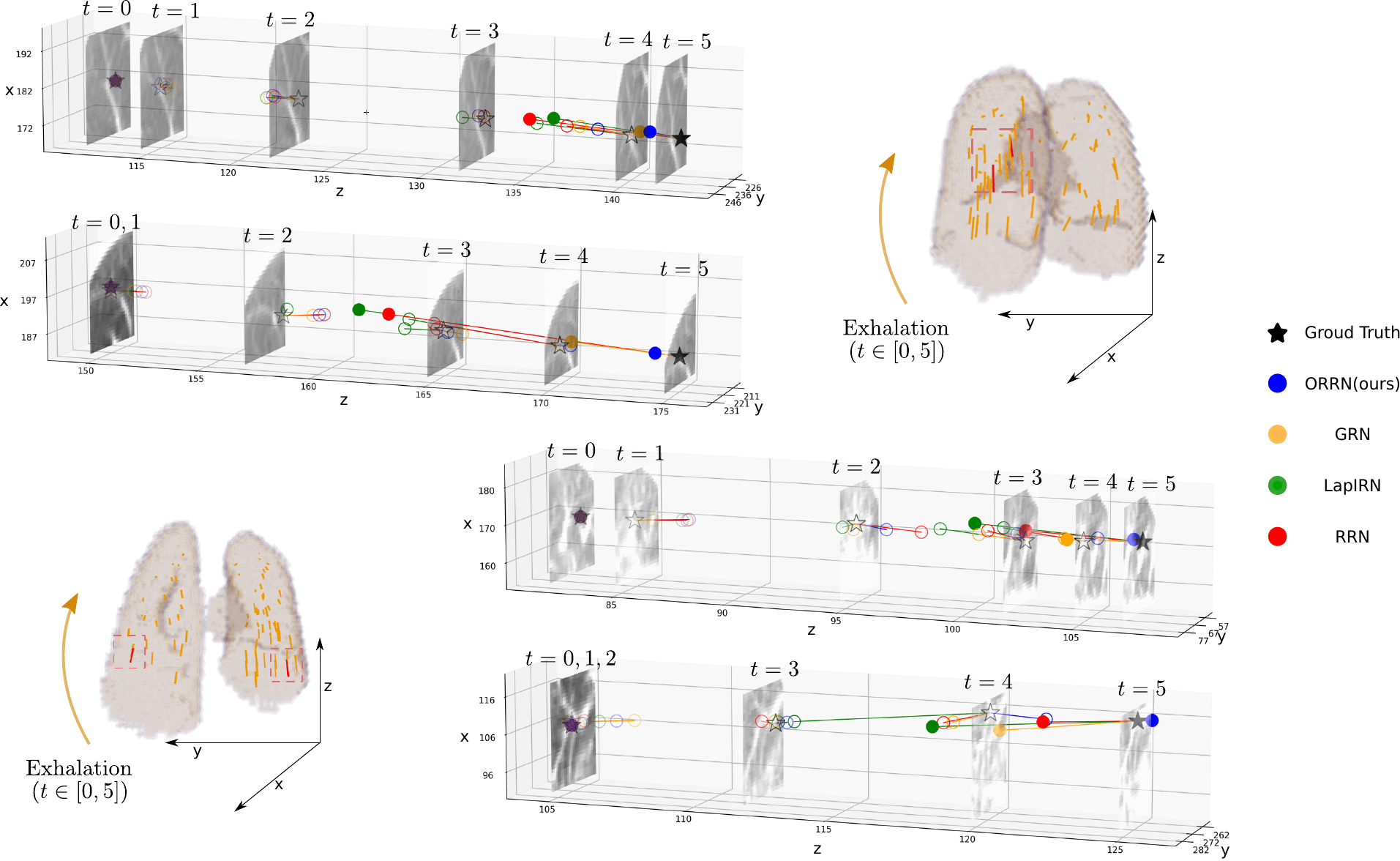}
\caption{A visualization of sampled landmarks' registration on DIR-Lab case 6 (top) and case 8 (bottom). Ground truth 4d landmarks are shown as $\star$ along with the axial image patches in the 4DCT where the \textcolor{modification_color}{landmarks are located}. Results for EE and EI phase scans are filled with color, whereas those for intermediate phases are hollow. From left to right on Z-axis (inferior to superior) shows respiratory motion from EI to EE scans. Lines are drawn from transformed landmarks to corresponding ground truth landmarks to illustrate TRE. Our method performs comparably to baseline methods where displacements are small and achieves higher accuracy for larger displacements}
\label{4dtracking}
\end{figure*}
\begin{figure}[!t]
\captionsetup{justification=centering,margin=0.1cm}
\centering
\includegraphics[width=0.46\textwidth]{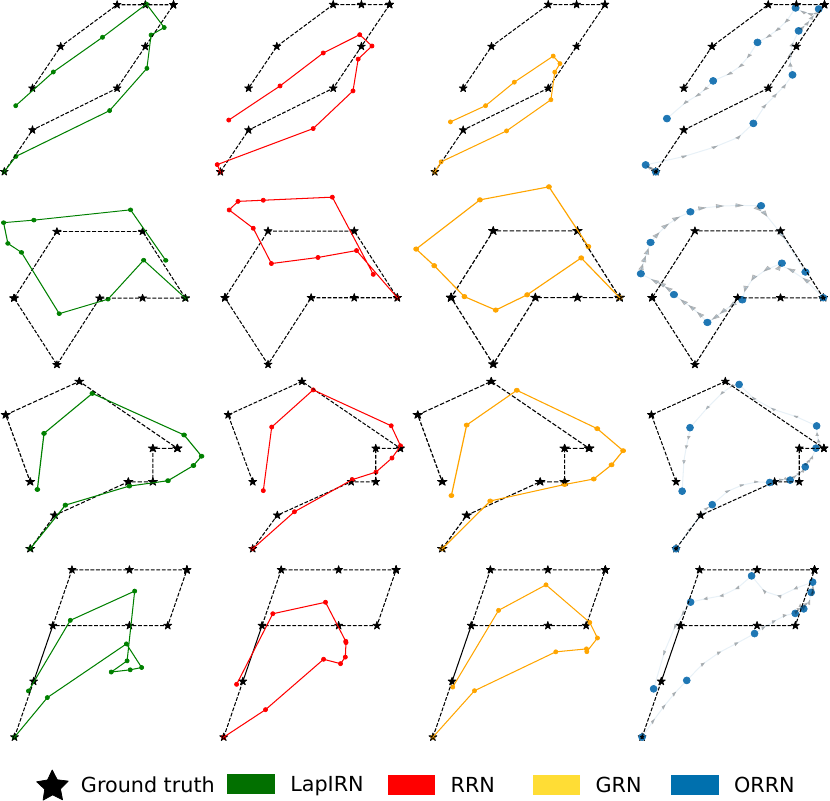}
\caption{Different methods' projected 2D trajectories on the coronal plane of sampled landmarks from CREATIS dataset. Intermediate velocity directions of landmarks are visualized with arrows for ORRN$^1$. In the above cases, landmarks' position and trajectory are better recovered by our method than those by baseline methods.}\label{4d_popi}

\end{figure}
\begin{figure*}[h!]
\includegraphics[width=0.95\linewidth]{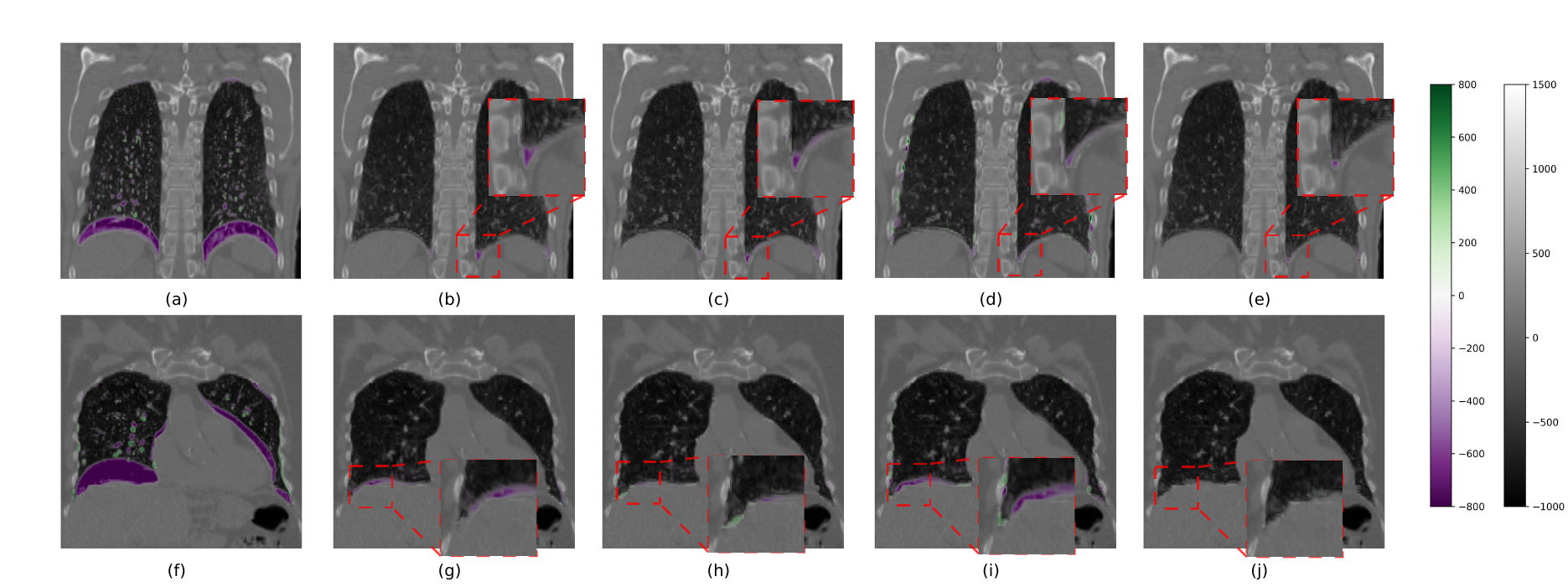}
\caption{Comparison of pair-wise registration performance of W/o registration (a, f), LapIRN (b, g), RRN (c, h), GRN (d, i), and ORRN$^2$ (e, j) of DIRLab 4DCT on two representative cases, case 2 (top) and case 10 (bottom). Each figure is a coronal overlay of image difference from the fixed image to a registered image onto the fixed image. Lung masks of EI phase images are used to filter out image difference outside the lung region. From the shown examples, ORRN$^2$ produces smaller image difference, especially around corners and edges of the lung which frequently are challenging for DIR methods.}\label{overlay}
\end{figure*}

\section{Experiments and results}\label{result_and_exp}
To evaluate the proposed method, we conduct separate experiments for the group-wise and pair-wise registration tasks, as described in Section \ref{sec:problem_formulation}. We name them as \textbf{Task A} and \textbf{Task B} respectively from now. In the following subsections, we describe training and testing data, evaluation metrics, baseline methods for comparison, and our implementation details. Then, we conduct experiments to evaluate the method's accuracy, plausibility, and computational efficiency. Ablation studies are done to verify the effectiveness of each design choice. For simplicity, we denote an ORRN that has a single-resolution velocity network as $\textbf{ORRN}^1$, and that has a multi-resolution velocity network as $\textbf{ORRN}^2$.

\subsection{Data}
\subsubsection{Training data}
We aggregate two publicly available 4DCT datasets as our taining dataset, namely the SPARE challenge\cite{Shieh2019} and 4D-Lung\cite{Hugo2016}, \cite{Hugo2017}, \cite{Balik_2013}, \cite{Roman2012}. Both dataset have phase-sorted 4DCT of lung. Specifically, the SPARE challenge dataset has a Monte Carlo subset that contains projections simulated from actual patient 4D CTs. We include the entire subset of 32 scans from 12 different patients. The 4D-Lung dataset is obtained from The Cancer Imaging Archive \cite{Clark_2013}. The dataset contains 4D fan-beam CT (FBCT) of 20 patients who have locally-advanced, non-small cell lung cancer. We select 41 scans out of all 4D FBCTs to avoid significant motion artifacts and an unbalanced distribution across patients. In total, we collected 73 4D CT scans. The dataset is split into a training and validation set with a size of 60 and 13 scans. Two sets do not contain scans from the same patient. To focus on the lung, we crop the lung region based on a lung mask generated with an automatic lung segmentation algorithm \cite{lungseg} on every EI phase image. Extra padding is added to the lung mask's boundaries to ensure lung regions are fully included. We perform image augmentation \cite{garcia_torchio_2021} including affine transformation, gaussian blurring, elastic deformation, and image contrast on the training data.

\subsubsection{Testing data}
We evaluated two separate 4DCT datasets that contain manually labeled anatomical landmarks, namely DIR-Lab 4DCT\cite{Castillo2009}, \cite{Castillo_2009} and CREATIS\cite{popi} dataset. The DIR-Lab 4DCT dataset contains ten 4DCT scans from different patients acquired as part of the radiotherapy planning process for the treatment of thoracic malignancies. Between the EI and EE phase image pairs, 300 landmarks were manually annotated. Additionally, 75 corresponding landmarks were identified at all phase images during exhalation (00-50\%). DIR-Lab 4DCT provides 3000 landmarks for evaluating registration between EI and EE pairs and 750 4D landmarks. The CREATIS dataset includes six phase-sorted 4DCT scans from different patients. For cases 1-3, it has 100 manually labeled landmarks on all 10 phase images. Case 4-6 contains around 100 landmarks at EI and EE phases. The dataset constitutes 600 landmarks for EI, EE pairs, and 300 4D landmarks. Among all testing cases, cases 6 and 8 in DIRLab 4DCT and case 2 in CREATIS dataset are cases with which the previous learning-based methods struggled due to their large and complex deformation. Similar to training data, the lung-cropping operation was applied to both testing datasets.

\subsection{Evaluation metrics}
\subsubsection{Target Registration Error}
To assess the accuracy of the proposed lung registration method, we compute the mean Target Registration Error (TRE)\cite{611354} using the annotated landmarks set as ground truth. 
For Task A, we have landmarks set $P^T = \{ \mathbf{p}_t |t = 0,\cdots,T-1\}$. Let $n$ be the size of an element in $P$. Mean 4D TRE of a set of displacement $\Phi^T$ estimated by ORRN is defined as:
\begin{equation}
    \label{eqn:tre4d}
    \text{mTRE}(\Phi^T) = \frac{1}{n(T-1)}\sum_{t \ne 0}\sum_{}\| \mathbf{p}_0  \oplus \Phi_t  - \mathbf{p}_t\|_2
\end{equation}
For Task B, landmarks set $P = \{\mathbf{p}_{f}, \mathbf{p}_{m}\}$ contains landmarks identified on the fixed and moving images. The mean TRE is defined as:
\begin{equation}
\begin{split}
    \label{eqn:trepw}
    \text{mTRE}(\Phi_1) = \frac{1}{n}\sum \|\mathbf{p}_{f} \oplus \Phi_1 -\mathbf{p}_{m} \|_2 \\
\end{split}
\end{equation}

\subsubsection{Jacobian Determinants of Displacement Fields}
To evaluate the plausibility of a predicted displacement field $\Phi$, we calculate its Jacobian determinants $\mathcal{J}_{\Phi}$\cite{DBLP:journals/corr/abs-2003-09514}. A value of 1 in $\mathcal{J}_{\Phi}$ indicates the underlying volume does not change. The volume undergoes expansion and compression at $\mathcal{J}_{\Phi}>1$, $\mathcal{J}_{\Phi}<1$ respectively. Image foldings occur where the value of $\mathcal{J}_{\Phi}$ is less than 0. We focus on the standard deviation of $\mathcal{J}_{\Phi}$, which is an indicator of the overall smoothness of the displacement field, and the fraction of negative value in Jacobian determinant that reflects the amount of unrealistic image foldings.

\mathchardef\mhyphen="2D
\begin{table*}[!ht]
\setlength\tabcolsep{1.3em}
\centering
\captionsetup{justification=centering,margin=0.5cm}
\caption{Target Registration Error of comparison methods on \textcolor{modification_color}{Task B}. 300 landmarks and 100 landmarks are evaluated for each scan in DIR-Lab and CREATIS datasets, respectively. Statistical significance is tested using a Wilcoxon Signed-rank Test on all available landmark results. Significance level are defined as: $^{*}$ for $p < 0.05$,  $^{**}$ for $p < 0.01$, $^{***}$ for $p < 0.001$}
\begin{adjustbox}{width=0.95\textwidth}
\begin{tabular}{c|c|cc|cccccccc|cc}
\toprule
\multicolumn{1}{c}{} &  \multicolumn{1}{c}{} & \multicolumn{2}{c}{Traditional method}  & \multicolumn{8}{c}{Unsupervised learning method}  & \multicolumn{2}{c}{Ours} \\ \cmidrule{3-4}\cmidrule{5-12}\cmidrule{13-14}
    & W/o reg        &Me11 & pTVreg   & MJ-CNN  & LRN  & MANet   & OSL  & Li22 & LapIRN  & RRN& GRN & ORRN$^1$ & ORRN$^2$ \\ \midrule
4DCT 01    & $3.89$  & $1.02$ & $0.77$     & $1.20$     & \boldmath$0.98$   & $1.06$       & $1.21$ & $1.13$  & $1.00$  & $0.99$ & \boldmath$0.98$ &  $0.99$ & $1.03$ \\
4DCT 02    & $4.34$   & $1.06$ &  $0.75$    & $1.13$     & \boldmath$0.98$   & $1.02$       & $1.13$ & $1.04$  & $1.00$  & $1.06$ & $0.99$ & $1.01$ & $0.99$\\
4DCT 03    & $6.94$   & $1.19$ &  $0.93$    & $1.30$     & \boldmath$1.14$   & $1.25$       & $1.32$ & $1.54$  & $1.21$  & $1.19$ & $1.19$ & $1.15$  & $1.18$\\
4DCT 04    & $9.83$   & $1.57$ &  $1.26$    & $1.55$     & \boldmath$1.39$            & $1.45$       & $1.84$ & $1.66$  & $1.45$  & $1.50$ & $1.45$ & $1.41$ & \boldmath$1.39$\\
4DCT 05    & $7.48$   & $1.73$ &  $1.07$    & $1.72$     & $1.43$            & $1.60$       & $1.80$ & $1.76$  & $1.50$  & $1.50$ & $1.43$ & $1.45$   &\boldmath$1.37$\\
4DCT 06    & $10.89$ & - &  $0.83$    & $2.02$     & $2.26$            & $2.00$       & $2.30$ & $1.90$  & $1.69$  & $1.63$ & $1.71$ & $1.42$& \boldmath$1.32$ \\
4DCT 07    & $11.03$  & - &  $0.80$    & $1.70$     & \boldmath$1.42$   & $1.56$       & $1.91$ & $1.78$  & $2.05$  & $2.17$ & $1.47$ &  $1.59$ & $1.44$\\
4DCT 08    & $15.0$   &- &  $1.01$    & $2.64$     & $3.13$            & $2.40$       & $3.47$ & $2.94$  & $3.13$  & $3.29$ & $1.85$ &  \boldmath$1.75$ & $1.81$\\
4DCT 09    & $7.92 $  &- &  $0.91$    & $1.51$     & $1.27$            & $1.46$       & $1.47$ & $1.74$  & $1.55$  & $1.49$ & $1.45$ &  $1.30$ & \boldmath$1.26$\\
4DCT 10    & $7.30$  &  -   &  $0.84$    & $1.79$     & $1.93$            & $1.52$       & $1.79$ & $1.70$  & $1.56$  & $1.52$ & $1.30$ &  $1.34$ & \boldmath$1.27$\\ \midrule
Avg        & $8.46$  & - &  $0.92$    & $1.66$     & $1.59$            & $1.53$       & $1.83$ & $1.71$  & $1.61$  & $1.63$ & $1.38$ &  $1.34$  & \boldmath$1.31$   \\ 
std        & $5.48$    & - &    -            &      -             &-           &-&-&-&                $2.10$ & $1.88$  & $1.23$ &$1.12$ & $1.20$\\
sig level &   ***    & -&   -            &      -             &     -      &-&-&-&***                 &***    &***& ***&\\ \bottomrule \toprule
CREATIS 01  & $5.90$ & - &  $0.80$    &      -      &    -     &                   -     & $1.09$ & $1.34$  & \boldmath$0.84$ & $0.94$ & $0.85$ & $0.90$ & $0.92$ \\
CREATIS 02  & $14.04$ &- &  $1.09$  &-&-&-                                               & $1.56$ & $1.74$  & $1.74$ & $1.75$ & $1.42$ & \boldmath$1.33$ & \boldmath$1.33$ \\
CREATIS 03  & $7.67$ &- &  $0.82$  &-&-&-                                               & $1.40$ & $1.577$ & $0.91$ & $1.15$ & $1.00$ & $0.94$  & \boldmath$0.90$\\
CREATIS 04  & $7.32$ & - &  -       &-&-&-                                               & $1.17$ & $1.64$  & $0.91$ & $1.19$ & $0.99$ & $0.92$ &  \boldmath$0.86$\\
CREATIS 05  & $7.09$ & - &  -       &-&-&   -                                            & $1.30$ & $1.26$  & $1.06$ & $1.17$ & \boldmath$1.03$ & $1.07$ & $1.04$\\
CREATIS 06  & $6.68$ & - & -        &-&-&-                                               & $1.27$ & $1.45$  & $0.96$ & $1.03$ & \boldmath$0.92$ & $1.01$ & $0.95$\\ \midrule
Avg & $8.08$   & -    & -        &-&-&-                                                 & $1.49$  & $1.50$  &$1.07$  & $1.14$ & $1.03$ & $1.03$&  \boldmath$1.00$ \\
std &   $5.61$  &   -   &          -&    -               &  -         &-&-&-&             $1.17$   &  $1.02$  &$0.82$& $0.78$ & $0.78$\\
Sig level &  ***   &       -&-         &  -                 &      -     &-&-&- & /                 &***    &/& / &\\
\bottomrule
\end{tabular}
\end{adjustbox}

\label{pairwise}
\end{table*}

\subsection{Implementation}\label{implementation}
A fixed-step Euler solver is used throughout the experiments. For group-wise registration (Task A), we assume that lung 4DCT represents a cyclic motion. Based on the assumption, we model respiratory motion in both forward and backward frame directions: (1) $[0\rightarrow 1\rightarrow  \cdots \rightarrow  5]$, (2) $[0\rightarrow9 \rightarrow  \cdots \rightarrow  5]$. For each direction, we use $T=6$, and from $t=0$ to $t=T$, we integrate ODE through five time intervals in a piece-wise manner by using the previously integrated displacement field as the initial condition for the following interval. For pair-wise registration (Task B), we integrate the ODE from  $t=0$ to $t=1$. During training, a random phase image is registered to the fixed image (EI) in the same 4DCT scan every iteration. The baseline methods are trained similarly. Relatively Large integration step sizes of $0.5$ and $0.2$ are used during training Task A and Task B to save training time and reduce memory requirements. To improve accuracy, we used a smaller step size of $0.25$ (A) and $0.1$ (B) during test time. The effect of varying step sizes at inference time is explored more in one of our ablation studies \ref{ablation_step_size}. 

Our ORRN is trained with a batch size of 1 and an Adam optimizer with an initial learning rate of 1e-4. We train ORRN$^1$ for 250 epochs for both tasks. ORRN$^2$ is trained from ORRN$^1$'s trained weight for 250 more epochs. For all experiments, CT images are resized to a spatial resolution of $192\times160\times192$. Then, their intensity is thresholded to  $[-1000, 500]$ HU and rescaled to $[0, 1]$ before being passed to the neural networks. In our experiments, ORRN$^1$ and ORRN$^2$ predict deformation at quarter and half of original CT resolution respectively. The proposed method is implemented in Pytorch version 1.8 \cite{NEURIPS2019_9015} with the Neural ODEs framework, tested on a Ubuntu 20.04 machine with an AMD Ryzen5 5600X CPU and a GeForce RTX 3090 GPU. During training, the proposed method requires 5.7 GB (ORRN$^1$) and 10.5 GB (ORRN$^2$) memory\footnote{GPU memory is measured as maximum memory allocated by PyTorch.} for Task A, and 12.9 GB for Task B. Our implementation will be available at \url{https://github.com/Lancial/orrn\_public}.

\subsection{Baselines}

The proposed method is compared to several recent DIR and respiratory motion estimation methods. These methods include non-learning methods (optimization-based and bio-mechanical model-based) such as pTVreg\cite{vishnevskiy2017isotropic}, Me11\cite{Metz2011}, LRME\cite{9220764}, and La21 \cite{Ladjal2021}. Unsupervised learning methods include MJ-CNN\cite{Jiang2020}, LungRegNet(LRN)\cite{Fu2020}, MANet\cite{Yang_2021}, OSL\cite{fechter2019}, Lung-CRNET\cite{Lu_2021}, RRN\cite{He2021}, LapIRN\cite{Mok2020}, and GRN\cite{Hansen2021}. While the above unsupervised learning methods were evaluated on the same data, their training data varied in size and diversity. Therefore, we trained RRN\footnote{\url{https://github.com/Novestars/Recursive_Refinement_Network}}, LapIRN\footnote{\url{https://github.com/cwmok/LapIRN}}, and GRN\footnote{\url{https://github.com/multimodallearning/graphregnet}} with our training data, utilizing their open-sourced implementation. For RRN, we use a three-level flow estimator as it has the best empirical performance. Its regularization loss is changed from total variation to our smoothness loss function. For GRN, we threshold CT images to $[-1000, 1500]$ Hounsfield Unit (HU) to be consistent with the original work because it relies on a fixed feature extractor.

\subsection{Evaluation on 4D data (Task A)}
Group-wise registration is challenging as multiple CT images are required to be registered simultaneously to a common space. Table \ref{4dtable} shows the comparison of 4D mean TRE produces by different methods on two evaluation datasets. Among all learning-based methods, ORRN$^1$ achieved the smallest mean TRE, which reduces initial error from 5.73mm, 6.28mm to 1.24mm, 1.23mm for two evaluation datasets. Our method improves registration accuracy significantly at more complex cases 6 and 8 in DIR-Lab and case 2 in CREATIS where other learning-based methods have difficulties registering accurately. With respect to the SOTA learning-based groupwise method LCRN, our method achieves 11\% decrease in TRE. Compared with non-learning methods, ours still has a gap to the SOTA conventional method LRME-4DCT\cite{9220764}. However, it is more accurate than a conventional nD+t B-splines group-wise registration method\cite{Metz2011}, and a previous biomechanical model-based method \cite{Ladjal2021}. Figure \ref{4dtracking} shows that ORRN$^1$ can track initial landmarks more accurately when they experience large motion amplitude than baseline methods. In addition, we show a set of sampled landmarks tracking results for the entire respiratory cycle in Figure \ref{4d_popi}. In contrast to baseline methods, ORRN$^1$ produces the most similar landmark trajectories to the manually labeled ground truth. Furthermore, using our ODE formulation, the proposed method can produce a piece-wise smooth trajectory.

\subsection{Evaluation on 3D data (Task B)}\label{3deval}
We show that the propose method can also serve as an general framework for DIR of image pairs. Table \ref{pairwise} reports the mean TRE of comparison methods on two evaluation datasets. Both ORRN$^1$ and ORRN$^2$ can outperform other unsupervised learning methods with smaller mean TRE. For DIRLab cases 6 and 8 and CREATIS case 2, which have large deformation, they show remarkable improvement over previous unsupervised learning methods. ORRN$^1$ has reasonable accuracy in other test cases, whereas ORRN$^2$ improves upon ORRN$^1$ over those cases with a multi-scale architecture. Overall, we are able to improve pair-wise registration accuracy of EE and EI image pairs by 5\%, 14\%, and 18\% over the previous SOTA unsupervised method GRN, MANet, and LRN, respectively. Figure \ref{overlay} shows a qualitative image registration result. ORRN$^2$ accurately registers EE and EI image pairs. It produces the slightest image differences on average and does notably better at labeled lung edge and corner areas. Admittedly, there are still gaps between the proposed method to SOTA conventional methods such as pTVreg\cite{vishnevskiy2017isotropic}. However, we believe ORRN still has space for improvement by further applying multi-resolution refinement in the velocity estimation module.

\subsection{Plausibility of Deformation}
A good deformation field should cause no or minimum image folding. For Task A, deformation plausibility is evaluated by computing the average fraction of negative values and standard deviation of Jacobian determinants of all displacement fields predicted for every moving image. Our method has an average of 0.0002\% of negative values and $0.13$ standard deviation. In contrast, baseline methods have RRN (0.003\%, 0.12), LapIRN (0.0\%, 0.17), and GRN (0.06\%, 0.19). For Task B, the average percentage of negative values and standard deviation of Jacobian Determinants of predicted displacement field at $t=1$ is $0\%$ and $0.14$ for ORRN$^1$, $0.0006\%$ and $0.25$ for ORRN$^2$ on two evaluation datasets. Our method has comparable plausibility to RRN (0.005\%, 0.14), LapIRN (0\%, 0.19), and GRN (0.1\%, 0.23). The plausibility evaluations for two tasks show that our method can predict well-regularized and plausible deformation.

\subsection{Efficiency}
We compare the computational efficiency of our ORRN and baseline methods in Table \ref{efficiency_table} using the metrics of neural networks' parameter size, inference-time GPU memory, and computation time. Our network does not have the fewest network parameters but has significantly fewer parameters than RRN (15M), where our architecture is modified from. Among four learning-based comparison methods, it requires the smallest GPU memory. In terms of computation speed, ORRN is a little slower than other baselines but faster than previous most accurate unsupervised method (GRN). \textcolor{modification_color}{Compared to other methods' speed from the literature, such as learning-based LRN methods: (60s), MANet (1s), MJ-CNN (1.4s), and traditional methods: LRME-4DCT (9.2s), pTVreg (130s), Me11 (1 hour), our ORRN leaves an impression of fast computation even though different computing devices were used.} In summary, our method has comparable efficiency to other efficient learning-based methods.

\begin{table}[h!]
\setlength\tabcolsep{0.7em}
\centering
\caption{Comparison of efficiency of different methods in terms of output format (along with resolution), GPU memory, network parameter size, and computation speed.}
\begin{adjustbox}{width=0.47\textwidth}
    \begin{tabular}{lccccc}
    \toprule
 \multirow{2}{*}{Methods} & Output & Memory  & \# Params & Time &\multirow{2}{*}{Step size} \\
  & format & (GB)  & (Million) & (s/phase) \\\midrule
RRN & dense, $1/4$ & $0.8$ & $15.0$ & $0.16$ &-\\
LapIRN & dense, full & $ 4.5$  &$0.9$ & $0.29$ &-\\
GRN & key points  & $3.5$  & $0.07$ & $1.45$ &-\\
A-ORRN$^1$ & dense, $1/4$ & $0.6$  & $1.46$ & $0.22$ & $0.25$\\
B-ORRN$^1$ & dense, $1/4$ & $0.5$  & $1.46$ & $0.53$ & $0.1$\\
B-ORRN$^2$ & dense, $1/2$ & $1.4$  & $1.58$ & $0.87$ & $0.1$\\

    \bottomrule
    \end{tabular}
\end{adjustbox}
\label{efficiency_table}
\end{table}

\subsection{Ablation Study}
\label{ablation_study}
We conduct several ablation studies to further explore our ORRN including integration step sizes, recursive strategies, and different architectural choices.

\begin{table}[b!]
\setlength\tabcolsep{0.5em}
\centering
\caption{\label{ablation_step_size} Effect of varying step sizes (S) or number of recursive steps (R) on the proposed ORRN(S) and its variant dRRN(R). The 1st column indicates related training parameters. Bad results (mTRE$ > 2.5$) are not neglected.}
\begin{adjustbox}{width=0.5\textwidth}
    \begin{tabular}{lccccc}
    \toprule
    & \multicolumn{5}{c}{Testing Scenarios}\\ \cmidrule{2-6}
     & $S=1$ & $S=0.5$ & $S=0.25$& $S=0.2$& $S=0.1$\\\cmidrule{1-6}
    A-ORRN$^1(1)$ & $1.31(1.00)$ & \boldmath$1.28(1.10)$& $1.30(0.99)$ & $1.30(0.99)$ & $1.31(0.99)$ \\ 
    A-ORRN$^1(0.5)$ & $1.89(1.72)$ & $1.25(0.96)$& \boldmath$1.24(0.95)$ & $1.24(0.95)$ & $1.24(0.95)$ \\
    B-ORRN$^1(1)$ & $2.10(2.86)$ & $1.65(1.82)$& $1.51(1.26)$ & $1.50(1.19)$ & \boldmath$1.50(1.10)$ \\ 
    B-ORRN$^1(0.5)$ & - & $1.48(1.48)$ & $1.42(1.31)$ & $1.41(1.27)$ & \boldmath$1.40(1.17)$ \\ 
    B-ORRN$^1(0.25)$ & - &- &$1.38(1.25)$ & $1.35(1.17)$ & \boldmath$1.34(1.11)$ \\ 
    B-ORRN$^1(0.2)$ & -& -& $1.33(1.21)$ & $1.30(1.13)$ & \boldmath$1.29(1.07)$ \\ 
    
 \midrule
    & $R=1$ & $R=2$ & $R=4$& $R=5$& $R=10$\\\cmidrule{1-6}
    A-dRRN$^1(1)$ & $1.31(1.00)$ & \boldmath$1.29(0.99)$ & $1.34(0.99)$ & $1.36(0.99)$ & $1.45(1.04)$ \\
    A-dRRN$^1(2)$ & $1.31(1.04)$ & \boldmath$1.25(0.98)$ & $1.26(0.97)$ & $1.27(0.97)$ & $1.29(0.98)$\\
    B-dRRN$^1(1)$ & $2.10(2.86)$ & \boldmath$1.58(1.44)$ & $1.62(1.35)$ & $1.71(1.64)$ & $2.31(4.25)$ \\
    B-dRRN$^1(2)$ &-  & $1.50(1.64)$ & \boldmath$1.40(1.15)$ & $1.43(1.12)$ & $1.69(1.80)$\\ 
    B-dRRN$^1(4)$ & - &-  & $1.41(1.40)$ & \boldmath$1.37(1.29)$ & $1.38(1.15)$ \\ 
    B-dRRN$^1(5)$ & - &  -& $1.36(1.21)$ & \boldmath$1.31(1.12)$& $1.32(1.04)$ \\
\midrule
    
    Time(s/phase) & $0.08$ & $0.13$ & $0.23$ & $0.28$ & $0.53$\\ 
    \bottomrule
    \end{tabular}
\end{adjustbox}
\label{stepsize}
\end{table}

\subsubsection{Integration Step Sizes}\label{ablation1}
We explore the impacts of the ODE solver's step sizes on ORRN. The first block of Table \ref{stepsize} shows the mean TRE of ORRN$^1$ with varying step sizes evaluated on two datasets together. In addition to training and testing with the same step sizes, we explore the effects of varying step sizes at test time. When two step sizes are the same, mean TRE reduces from 2.10mm to 1.30mm as step sizes decrease from $1$ to $0.2$, demonstrating our recursive strategy's effectiveness. Additionally, in Task B, the mean TRE is reduced when the testing step-size becomes smaller without further training. The reduction is significant when the training step size is large, whereas it becomes less significant when training step sizes are small. In Task A, this is not observed when a step size becomes less than 0.25, probably because the deformation between consecutive phase images is too small for further decomposition to be helpful.

\begin{figure}[!t]
\captionsetup{margin=0.1cm}
\centering
\includegraphics[width=0.43\textwidth]{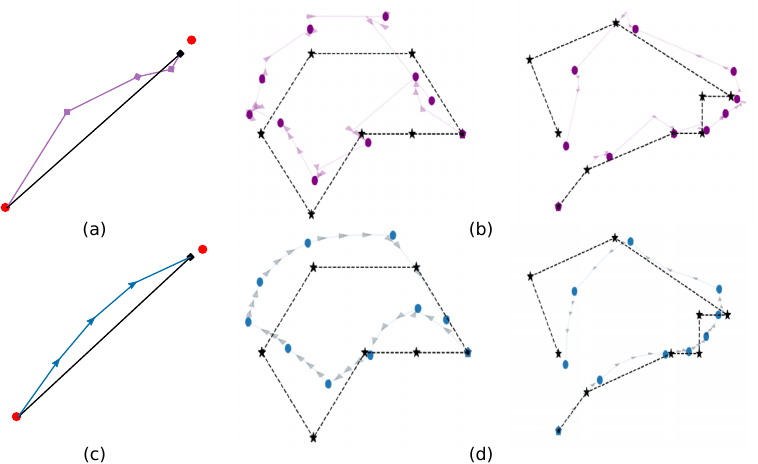}

\caption{Landmark trajectories visualization. (a), (c): idea visualization of a dRRN and ORRN. (b), (d): real landmarks trajectories in CREATIS dataset obtained by a dRRN and ORRN.}
\label{orrn_vs_recursive}
\end{figure}

\subsubsection{Recursive Strategies}\label{ablation2}
 We build a variant recursive network called \textbf{dRRN} that recursively predicts small displacement fields for composition for a number of recursions instead of predicting velocity. The variant has the same network architecture as ORRN. The same experiment as the previous ablation study is conducted for dRRN at the second block in Table \ref{stepsize}. The variant has a comparable mean TRE to the ORRN when the same number of steps is used for training and testing. However, in both Task A and B, dRRN's accuracy first improves a little but then degrades when the number of recursions increases at test time. This is because it becomes less stable when more recursions are used than in its training scenario. In Task A, two recursive registration methods also output different kinds of voxel trajectories, as shown in Figure \ref{orrn_vs_recursive}. Our ORRN produces more regular trajectories between consecutive 4DCT images than dRRN. Even though we don't explicitly regularize temporal smoothness, by studying previous work that emphasizes temporal trajectory smoothness, such as in \cite{Metz2011}, \cite{popi}, we believe that ORRN produces a similar velocity profile. One of our future directions is to impose explicit temporal regularization on this work.

\mathchardef\mhyphen="2D

\begin{table}[h!]
\setlength\tabcolsep{1em}
\centering
\caption{Experiments on different network design choices evaluated on DIR-Lab and CREATIS datasets. Wilcoxon Signed-Rank Tests are conducted between the best and other variants with all available landmarks. A $p < 0.001$ is indicated with $^{***}$.}\label{ablation3}
\begin{adjustbox}{width=0.47\textwidth}
\begin{tabular}{lll|ll}
\toprule
 & \multicolumn{2}{c}{Group-wise (Task A)}  &  \multicolumn{2}{c}{Pair-wise (Task B)}\\\cmidrule{2-3}\cmidrule{4-5}
 & TRE & Folds (\%)& TRE & Folds (\%)\\ \midrule
NoCorrV   & $1.31(1.03)^{***}$ & $0$ & $1.43(1.31)^{***}$  & $2^{-5}$\\ 
NoGRU  & $1.26(0.98)^{***}$ &  $0$  & - & - \\ 
ORRN$^1$  &\boldmath$1.24(0.95)$ &  $2^{-4}$ & $1.29(1.07)^{***}$ & $4^{-5}$  \\ 
ORRN$^2$    & - & - & \boldmath$1.26(1.15)$ & $6^{-4}$ \\ 
\bottomrule
\end{tabular}
\end{adjustbox}

\end{table}

\subsubsection{Network Architectures}
Lastly, we evaluate the influence of different architectural choices on registration quality. We construct four variants of ORRN: (1) \textit{NoCorrV}: cost volume is not computed, and moving features are directly used as input. (2) \textit{NoGRU}: Temporal feature encoding is not used (not evaluated for pair-wise registration), (3) ORRN$^1$:  a single-level velocity network, (4) ORRN$^2$: a multi-resolution velocity network. We report the mean TRE and the fraction of foldings for the variants evaluated with two test datasets in Table \ref{ablation3}. The results have confirmed the effectiveness of each of our \textcolor{modification_color}{design choices.}

\section{Discussion}\label{sec:discussion}
In this paper, we present a novel ODE-based Recursive Registration Network (ORRN) for learning accurate dense deformation in 4D lung CT. For group-wise registration (Task A), our method produces a trajectory for every voxel in a CT volume through an ODE integration. It yields the smallest mean TRE at every discrete phase in 4DCT scans among all learning-based comparison methods. Our proposed method is also an accurate framework for pair-wise registration \textcolor{modification_color}{(Task B)}. It surpasses the previous SOTA unsupervised learning methods for pair-wise registration between EE and EI images on two public datasets. Besides accuracy, we have shown that the proposed ORRN generates reasonably regularized deformation and has comparable computation efficiency to current SOTA learning-based DIR methods. \textcolor{modification_color}{Comparing to traditional methods, ORRN out-perform a previous group-wise registration method \cite{Metz2011} but still doesn't surpass the state-of-the-art methods \cite{vishnevskiy2017isotropic, 9220764}. In practical applications, traditional methods can be used for optimizing registration accuracy with a long computation duration. Whereas in applications that require fast computation for a large number of images, our method could be chosen given its accuracy and efficiency.}

While other architectural choices, such as cost volume, and Convolutional GRU, have proven to be effective in our ablation studies, the ODE-based recursive strategy has played the most vital role in reducing average registration error (from 2.10 mm to 1.30 mm). To alleviate the computational burden that arises when training with small integration step sizes, our approach can be trained using relatively large step sizes but still attain high accuracy by adjusting to smaller step sizes during inference. For example, ORRN$^1$ with an integration step size of 1 has GPU memory usage of 3.0 GB during training and can reduce mean TRE from 8.40mm to 1.50mm on two testing datasets in the pair-wise task by using a step size of 0.1 during inference.

A common limitation of learning-based DIR methods is the enormous computational burden when processing volumetric data. The proposed method also requires relatively large GPU memory when it uses small step sizes during training. One solution is to use more lightweight network architecture. Given the decomposition ability of the recursive strategy, it is possible to maintain the current accuracy while simplifying the network architecture. Another potential solution is taking advantage of Neural ODE's adjoint sensitive method, which has a constant memory requirement\cite{Chen2018}. This work does not use the adjoint method because it makes training unstable towards the end, likely due to large step sizes. While this work mainly focuses on predicting accurate dense displacement fields that align 4D CT images, we think another direction is to focus on introducing existing or new temporal regularization loss such as the cyclic loss in \cite{fechter2019} and temporal smoothness constraints in \cite{Metz2011}, \cite{popi} to further regularize the predict voxel trajectory. In the end, we believe that ORRN can be extended to deformation estimation in other volumetric medical image data such as those of different imaging sources (MRI, X-ray etc.) and different organs (heart, liver, brain etc.).

\section{Conclusion}

The proposed method accurately estimates deformation presents in lung 4DCT through pair-wise and group-wise DIR. It reaches average target registration errors of 1.24mm and 1.26mm in two types of DIR tasks, respectively, which are 11\% and 6\% less than the errors produced by previous unsupervised state-of-the-art methods from existing literature \cite{Lu_2021}, \cite{Hansen2021}. The computation time required to register a phase image using this new method is less than one second, making it faster than traditional methods and similar to other efficient learning-based methods. Our future work will focus on imposing temporal regularization to the proposed method for more realistic trajectory prediction, and applying it to challenging clinical settings such as automated robotic applications, such as lung biopsy needle insertion \cite{Schreiber_2022}, and surgical tool manipulation \cite{Jingbin_2021}.

%


\section*{APPENDIX}
\subsection{Network Architecture Details}
Table \ref{netdetails} shows the detailed network architecture of ORRN. A kernel size of 3 and a padding of 1 is used for all 3D convolutional layers. We use \textit{LeakyReLU} as our activation function except for the Convolutional GRU.
\begin{table}[h!]
\setlength\tabcolsep{0.5em}
\centering
\caption{Detailed network architecture of ORRN.}\label{netdetails}
\begin{adjustbox}{width=0.48\textwidth}
    \begin{tabular}{cc|cccc}
    \toprule
 &  &  In:Out  & \multirow{2}{*}{Stride} & Output & \multirow{2}{*}{Skip}  \\
 &  &  Channels  &  & Dim  &   \\ \midrule
 & \#1 & 1:16 & 2 & 1/2 & -\\
 & \#2 & 16:16 & 1 & 1/2 & -\\
 Feature& \#3 & 16:16 & 1 & 1/2& (\#15)$^*$ \\
 Pyramid& \#4 & 16:32 & 2 & 1/4& -\\
 & \#5 & 32:32 & 1 & 1/4& -\\
 & \#6 & 32:32 & 1 & 1/4& \#10\\
 \midrule
\multirow{3}{*}{ConvGRU}  & Update \#7 & 64:32 & 1 & 1/4 & -\\
   & Reset \#8 & 64:32 & 1 & 1/4 & -\\
 & New \#9& 64:32 & 1 & 1/4 & -\\
 \midrule
 
 & \#10 & 224:64 & 1 & 1/4 & \#11-13\\
 & \#11 & 288:48 & 1 & 1/4 & \#12-13\\
$\textbf{VN}^{1/4}$ & \#12& 336:32 & 1 & 1/4 & \#13\\
 & \#13& 368:16 & 1 & 1/4 & (\#15)\\
 & \#14& 16:3 & 1 & 1/4 & -\\
  \midrule
 & \#15 & 78:32 & 1 & 1/2 & \#16\\
  $\textbf{VN}^{1/2}$ & \#16 & 110:16 & 1 & 1/2 & -\\
 & \#17& 16:3 & 1 & 1/2 & \\

\bottomrule
\multicolumn{6}{l}{\footnotesize $^*$The parenthesis indicates that a skip exists only if a multi-resolution velocity network is used.} 

    \end{tabular}
\end{adjustbox}

\end{table}

\subsection{Local Cost Volume Computation}\label{localcostvolume}
A cost volume contains the correspondence for voxels with other voxels within a defined search space. Let us denote a warped feature map and a fixed feature map as $\mathcal{F}_w, \mathcal{F}_f \in \mathbb{R}^{H\times W \times D \times C}$. A local cost volume that only computes the correlation of voxels on $\mathcal{F}^w$ with its cubical neighborhoods with radius $r$ is defined as $\mathbb{CV}_(\mathcal{F}_w, \mathcal{F}_f) \in \mathbb{R}^{H\times W \times D \times (2r+1)}$ . Each of its entries can be computed as:
\begin{equation}
\begin{aligned}
    \small
    \mathbb{CV}_{p, k}(\mathcal{F}_w, \mathcal{F}_f) & = \sum_{c}\frac{\mathcal{F}_w(p, c) - \mu}{\sigma} \times \frac{\mathcal{F}_f(p+k, c) - \mu}{\sigma}\\
\end{aligned}
\end{equation}
where $c$ indicates a feature dimension index,  $p$ is a voxel position, and $k \in [-r, r]^3$ is position shift. A new index $p+k$ is an abbreviation for $(x+k_1, y+k_2, z+k_3)$. Similar to \cite{He2021} and \cite{uflow}, cost volume normalization is performed by with terms $\mu, \sigma$, which are the average and standard deviation of both feature maps $\mathcal{F}_w$ and $\mathcal{F}_f$ over their spatial and feature dimensions.

\subsection{Compose \& Scale}\label{composeandscale}
A vector field $v$ predicted by our proposed velocity network is a displacement field defined on a uniform Euler grid. It maps a warped feature map $\mathcal{F}_w$ to a fixed feature map $\mathcal{F}_f$. Let $t$ and $tm$ be the current integration time point and the time point related to a moving image respectively. To describe the motion of voxels that corresponds to the fixed image, we compose the previous voxel displacement $\Phi_t$ with $v$ to obtain $\Phi_{tm}'$ (an current estimate of voxel displacement from the fixed to the moving feature map). Lastly, we subtract $\Phi_t$ from $\Phi_{tm}'$ and scale the result based on difference between $t$ and $tm$ to obtain an estimation of voxel velocity:
\begin{equation}
\begin{aligned}
    \Phi_{tm}' =\ & \Phi_t \circ  v  + v \\
    \dot \Phi_t =\ & \frac{  \Phi_{tm}'  - \Phi_t }{t_m -t }. \\
\end{aligned}
\end{equation}

\bibliographystyle{IEEEtran}
\bibliography{reference} 

\begin{thebibliography}{10}
\providecommand{\url}[1]{#1}
\csname url@samestyle\endcsname
\providecommand{\newblock}{\relax}
\providecommand{\bibinfo}[2]{#2}
\providecommand{\BIBentrySTDinterwordspacing}{\spaceskip=0pt\relax}
\providecommand{\BIBentryALTinterwordstretchfactor}{4}
\providecommand{\BIBentryALTinterwordspacing}{\spaceskip=\fontdimen2\font plus
\BIBentryALTinterwordstretchfactor\fontdimen3\font minus
  \fontdimen4\font\relax}
\providecommand{\BIBforeignlanguage}[2]{{%
\expandafter\ifx\csname l@#1\endcsname\relax
\typeout{** WARNING: IEEEtran.bst: No hyphenation pattern has been}%
\typeout{** loaded for the language `#1'. Using the pattern for}%
\typeout{** the default language instead.}%
\else
\language=\csname l@#1\endcsname
\fi
#2}}
\providecommand{\BIBdecl}{\relax}
\BIBdecl

\bibitem{doi:10.3109/0284186X.2013.813638}
E.~S. Worm \emph{et~al.}, ``Variations in magnitude and directionality of
  respiratory target motion throughout full treatment courses of stereotactic
  body radiotherapy for tumors in the liver,'' \emph{Acta Oncologica}, vol.~52,
  no.~7, pp. 1437--1444, 2013, pMID: 23879645.

\bibitem{yang2008}
D.~Yang \emph{et~al.}, ``4d-ct motion estimation using deformable image
  registration and 5d respiratory motion modeling,'' \emph{Medical physics},
  vol.~35, pp. 4577--90, 11 2008.

\bibitem{Flampouri_2006}
S.~Flampouri \emph{et~al.}, ``Estimation of the delivered patient dose in lung
  imrt treatment based on deformable registration of 4d-ct data and monte carlo
  simulations,'' \emph{Physics in Medicine \& Biology}, vol.~51, no.~11, p.
  2763, may 2006.

\bibitem{10.1371/journal.pone.0271064}
J.~D. Tascón-Vidarte \emph{et~al.}, ``Accuracy and consistency of
  intensity-based deformable image registration in 4dct for tumor motion
  estimation in liver radiotherapy planning,'' \emph{PLOS ONE}, vol.~17, no.~7,
  pp. 1--15, 07 2022.

\bibitem{Hansen2021}
L.~Hansen and M.~P. Heinrich, ``{GraphRegNet}: Deep graph regularisation
  networks on sparse keypoints for dense registration of 3d lung {CTs},''
  \emph{{IEEE} Transactions on Medical Imaging}, vol.~40, no.~9, pp.
  2246--2257, sep 2021.

\bibitem{HERING2021102139}
A.~Hering \emph{et~al.}, ``Cnn-based lung ct registration with multiple
  anatomical constraints,'' \emph{Medical Image Analysis}, vol.~72, p. 102139,
  2021.

\bibitem{Zhang_2021}
Y.~Zhang \emph{et~al.}, ``Groupregnet: a groupwise one-shot deep learning-based
  4d image registration method,'' \emph{Physics in Medicine \& Biology},
  vol.~66, no.~4, p. 045030, feb 2021.

\bibitem{Lu_2021}
J.~Lu \emph{et~al.}, ``Lung-{CRNet}: A convolutional recurrent neural network
  for lung 4dct image registration,'' \emph{Medical Physics}, vol.~48, no.~12,
  pp. 7900--7912, nov 2021.

\bibitem{Fu2020a}
Y.~Fu \emph{et~al.}, ``Deep learning in medical image registration: a review,''
  \emph{Physics in Medicine {\&} Biology}, vol.~65, no.~20, p. 20TR01, oct
  2020.

\bibitem{Sotiras2013}
A.~Sotiras, C.~Davatzikos, and N.~Paragios, ``Deformable medical image
  registration: A survey,'' \emph{{IEEE} Transactions on Medical Imaging},
  vol.~32, no.~7, pp. 1153--1190, jul 2013.

\bibitem{VIERGEVER2016140}
M.~A. Viergever \emph{et~al.}, ``A survey of medical image registration –
  under review,'' \emph{Medical Image Analysis}, vol.~33, pp. 140--144, 2016,
  20th anniversary of the Medical Image Analysis journal (MedIA).

\bibitem{ffd2}
J.~Kybic and M.~Unser, ``Fast parametric elastic image registration,''
  \emph{IEEE Transactions on Image Processing}, vol.~12, no.~11, pp.
  1427--1442, 2003.

\bibitem{elastic2}
R.~Bajcsy and S.~Kovačič, ``Multiresolution elastic matching,''
  \emph{Computer Vision, Graphics, and Image Processing}, vol.~46, no.~1, pp.
  1--21, 1989.

\bibitem{demon}
J.-P. Thirion, ``Image matching as a diffusion process: an analogy with
  maxwell's demons,'' \emph{Medical Image Analysis}, vol.~2, no.~3, pp.
  243--260, 1998.

\bibitem{Ladjal2021}
H.~Ladjal \emph{et~al.}, ``Towards non-invasive lung tumor tracking based on
  patient specific model of respiratory system,'' \emph{{IEEE} Transactions on
  Biomedical Engineering}, vol.~68, no.~9, pp. 2730--2740, sep 2021.

\bibitem{Fei_2021_ICRA}
F.~Liu \emph{et~al.}, ``Real-to-sim registration of deformable soft tissue with
  position-based dynamics for surgical robot autonomy,'' in \emph{2021 IEEE
  International Conference on Robotics and Automation (ICRA)}, 2021, pp.
  12\,328--12\,334.

\bibitem{popi}
J.~Vandemeulebroucke \emph{et~al.}, ``Spatio-temporal motion estimation for
  respiratory-correlated imaging of the lungs,'' \emph{Medical physics},
  vol.~38, pp. 166--78, 01 2011.

\bibitem{Metz2011}
C.~Metz \emph{et~al.}, ``Nonrigid registration of dynamic medical imaging data
  using nd+t b-splines and a groupwise optimization approach,'' \emph{Medical
  Image Analysis}, vol.~15, no.~2, pp. 238--249, apr 2011.

\bibitem{8360027}
I.~Y. Ha \emph{et~al.}, ``Model-based sparse-to-dense image registration for
  realtime respiratory motion estimation in image-guided interventions,''
  \emph{IEEE Transactions on Biomedical Engineering}, vol.~66, no.~2, pp.
  302--310, 2019.

\bibitem{9220764}
P.~Xue \emph{et~al.}, ``Lung respiratory motion estimation based on fast kalman
  filtering and 4d ct image registration,'' \emph{IEEE Journal of Biomedical
  and Health Informatics}, vol.~25, no.~6, pp. 2007--2017, 2021.

\bibitem{Song2017ARO}
G.~Song \emph{et~al.}, ``A review on medical image registration as an
  optimization problem,'' \emph{Current Medical Imaging Reviews}, vol.~13, pp.
  274 -- 283, 2017.

\bibitem{8902170}
K.~A.~J. Eppenhof \emph{et~al.}, ``Progressively trained convolutional neural
  networks for deformable image registration,'' \emph{IEEE Transactions on
  Medical Imaging}, vol.~39, no.~5, pp. 1594--1604, 2020.

\bibitem{gdlfire}
T.~Sentker, F.~Madesta, and R.~Werner, ``Gdl-fire: Deep learning-based fast 4d
  ct image registration,'' in \emph{Medical Image Computing and Computer
  Assisted Intervention -- MICCAI 2018}, A.~F. Frangi \emph{et~al.}, Eds.\hskip
  1em plus 0.5em minus 0.4em\relax Cham: Springer International Publishing,
  2018, pp. 765--773.

\bibitem{modelbasedartificial}
H.~Sokooti \emph{et~al.}, ``3d convolutional neural networks image registration
  based on efficient supervised learning from artificial deformations,''
  \emph{arXiv}, 2019.

\bibitem{Fu2020}
Y.~Fu \emph{et~al.}, ``{LungRegNet}: An unsupervised deformable image
  registration method for 4d-{CT} lung,'' \emph{Medical Physics}, vol.~47,
  no.~4, pp. 1763--1774, feb 2020.

\bibitem{spatialtransformer}
M.~Jaderberg \emph{et~al.}, ``Spatial transformer networks,'' \emph{CoRR}, vol.
  abs/1506.02025, 2015.

\bibitem{Balakrishnan2018}
G.~Balakrishnan \emph{et~al.}, ``Voxelmorph: A learning framework for
  deformable medical image registration,'' \emph{IEEE Transactions on Medical
  Imaging}, vol.~38, no.~8, pp. 1788--1800, 2019.

\bibitem{mlvirnet}
A.~Hering, B.~van Ginneken, and S.~Heldmann, ``mlvirnet: Multilevel variational
  image registration network,'' \emph{CoRR}, vol. abs/1909.10084, 2019.

\bibitem{cueaware}
X.~Cao \emph{et~al.}, ``Deformable image registration using a cue-aware deep
  regression network,'' \emph{IEEE Transactions on Biomedical Engineering},
  vol.~65, no.~9, pp. 1900--1911, 2018.

\bibitem{Yang_2021}
J.~Yang \emph{et~al.}, ``An unsupervised multi-scale framework with
  attention-based network ({MANet}) for lung 4d-{CT} registration,''
  \emph{Physics in Medicine \& Biology}, vol.~66, no.~13, p. 135008, jun 2021.

\bibitem{recurrentregistration}
R.~Sandk{\"{u}}hler \emph{et~al.}, ``Recurrent registration neural networks for
  deformable image registration,'' \emph{CoRR}, vol. abs/1906.09988, 2019.

\bibitem{recursive2}
B.~Hu \emph{et~al.}, ``Recursive decomposition network for deformable image
  registration,'' \emph{IEEE Journal of Biomedical and Health Informatics},
  vol.~26, no.~10, pp. 5130--5141, 2022.

\bibitem{Mok2020}
T.~C. Mok and A.~C. Chung, ``Fast symmetric diffeomorphic image registration
  with convolutional neural networks,'' in \emph{2020 IEEE/CVF Conference on
  Computer Vision and Pattern Recognition (CVPR)}, 2020, pp. 4643--4652.

\bibitem{Zhao2019}
S.~Zhao \emph{et~al.}, ``Recursive cascaded networks for unsupervised medical
  image registration,'' \emph{IEEE International Conference on Computer Vision
  (ICCV), 2019, pp. 10600-10610}, Jul. 2019.

\bibitem{fechter2019}
T.~Fechter and D.~Baltas, ``One-shot learning for deformable medical image
  registration and periodic motion tracking,'' \emph{IEEE Transactions on
  Medical Imaging}, vol.~39, no.~7, pp. 2506--2517, 2020.

\bibitem{Chi_2022}
W.~Chi, Z.~Xiang, and F.~Guo, ``Few-shot learning for deformable image
  registration in 4dct images,'' \emph{The British Journal of Radiology},
  vol.~95, no. 1129, jan 2022.

\bibitem{Jiang2020}
Z.~Jiang \emph{et~al.}, ``A multi-scale framework with unsupervised joint
  training of convolutional neural networks for pulmonary deformable image
  registration,'' \emph{Physics in Medicine {\&} Biology}, vol.~65, no.~1, p.
  015011, jan 2020.

\bibitem{He2021}
X.~He \emph{et~al.}, ``Recursive refinement network for deformable lung
  registration between exhale and inhale ct scans,'' \emph{arXiv}, Jun. 2021.

\bibitem{recursive1}
J.-Q. Zheng \emph{et~al.}, ``Recursive deformable image registration network
  with mutual attention,'' in \emph{Medical Image Understanding and Analysis},
  G.~Yang \emph{et~al.}, Eds.\hskip 1em plus 0.5em minus 0.4em\relax Cham:
  Springer International Publishing, 2022, pp. 75--86.

\bibitem{Chen2018}
R.~T.~Q. Chen \emph{et~al.}, ``Neural ordinary differential equations,''
  \emph{Advances in Neural Information Processing Systems}, 2018.

\bibitem{Niemeyer2019ICCV}
M.~Niemeyer \emph{et~al.}, ``Occupancy flow: 4d reconstruction by learning
  particle dynamics,'' in \emph{International Conference on Computer Vision},
  Oct. 2019.

\bibitem{Xu2021}
J.~Xu \emph{et~al.}, ``Multi-scale neural {ODEs} for 3d medical image
  registration,'' in \emph{Medical Image Computing and Computer Assisted
  Intervention {\textendash} {MICCAI} 2021}.\hskip 1em plus 0.5em minus
  0.4em\relax Springer International Publishing, 2021, pp. 213--223.

\bibitem{wu2021}
Y.~Wu \emph{et~al.}, ``Nodeo: A neural ordinary differential equation based
  optimization framework for deformable image registration,'' in
  \emph{Proceedings of the IEEE/CVF Conference on Computer Vision and Pattern
  Recognition (CVPR)}, June 2022, pp. 20\,804--20\,813.

\bibitem{cgru}
N.~Ballas \emph{et~al.}, ``Delving deeper into convolutional networks for
  learning video representations,'' \emph{arXiv}, 2015.

\bibitem{uflow}
R.~Jonschkowski \emph{et~al.}, ``What matters in unsupervised optical flow,''
  in \emph{Computer Vision – ECCV 2020: 16th European Conference, Glasgow,
  UK, August 23–28, 2020, Proceedings, Part II}.\hskip 1em plus 0.5em minus
  0.4em\relax Berlin, Heidelberg: Springer-Verlag, 2020, p. 557–572.

\bibitem{AVANTS200826}
B.~Avants \emph{et~al.}, ``Symmetric diffeomorphic image registration with
  cross-correlation: Evaluating automated labeling of elderly and
  neurodegenerative brain,'' \emph{Medical Image Analysis}, vol.~12, no.~1, pp.
  26--41, 2008, special Issue on The Third International Workshop on Biomedical
  Image Registration – WBIR 2006.

\bibitem{Shieh2019}
C.-C. Shieh \emph{et~al.}, ``{SPARE}: Sparse-view reconstruction challenge for
  4d cone-beam {CT} from a 1-min scan,'' \emph{Medical Physics}, vol.~46,
  no.~9, pp. 3799--3811, jul 2019.

\bibitem{Hugo2016}
G.~D. Hugo \emph{et~al.}, ``Data from 4d lung imaging of nsclc patients,''
  \emph{The Cancer Imaging Archive}, 2016.

\bibitem{Hugo2017}
G.~Hugo \emph{et~al.}, ``A longitudinal four-dimensional computed tomography
  and cone beam computed tomography dataset for image-guided radiation therapy
  research in lung cancer,'' \emph{Medical Physics}, vol.~44, no.~2, pp.
  762--771, 2017.

\bibitem{Balik_2013}
S.~Balik \emph{et~al.}, ``Evaluation of 4-dimensional computed tomography
  to~4-dimensional cone-beam computed tomography deformable image registration
  for lung cancer adaptive~radiation therapy,'' \emph{International Journal of
  Radiation Oncology Biology Physics}, vol.~86, no.~2, pp. 372--379, jun 2013.

\bibitem{Roman2012}
N.~O. Roman \emph{et~al.}, ``Interfractional positional variability of fiducial
  markers and primary tumors in locally advanced non-small-cell lung cancer
  during audiovisual biofeedback radiotherapy,'' \emph{International Journal of
  Radiation Oncology Biology Physics}, vol.~83, no.~5, pp. 1566--1572, aug
  2012.

\bibitem{Clark_2013}
K.~Clark \emph{et~al.}, ``The cancer imaging archive ({TCIA}): Maintaining and
  operating a public information repository,'' \emph{Journal of Digital
  Imaging}, vol.~26, no.~6, pp. 1045--1057, jul 2013.

\bibitem{lungseg}
J.~Hofmanninger \emph{et~al.}, ``Automatic lung segmentation in routine imaging
  is primarily a data diversity problem, not a methodology problem,''
  \emph{European Radiology Experimental}, vol.~4, no.~1, p.~50, 2020.

\bibitem{garcia_torchio_2021}
F.~P{\'e}rez-Garc{\'i}a, R.~Sparks, and S.~Ourselin, ``Torchio: a python
  library for efficient loading, preprocessing, augmentation and patch-based
  sampling of medical images in deep learning,'' \emph{Computer Methods and
  Programs in Biomedicine}, p. 106236, 2021.

\bibitem{Castillo2009}
R.~Castillo \emph{et~al.}, ``A framework for evaluation of deformable image
  registration spatial accuracy using large landmark point sets,''
  \emph{Physics in Medicine and Biology}, vol.~54, no.~7, pp. 1849--1870, mar
  2009.

\bibitem{Castillo_2009}
E.~Castillo \emph{et~al.}, ``Four-dimensional deformable image registration
  using trajectory modeling,'' \emph{Physics in Medicine and Biology}, vol.~55,
  no.~1, pp. 305--327, dec 2009.

\bibitem{611354}
C.~Maurer \emph{et~al.}, ``Registration of head volume images using implantable
  fiducial markers,'' \emph{IEEE Transactions on Medical Imaging}, vol.~16,
  no.~4, pp. 447--462, 1997.

\bibitem{DBLP:journals/corr/abs-2003-09514}
T.~C.~W. Mok and A.~C.~S. Chung, ``Fast symmetric diffeomorphic image
  registration with convolutional neural networks,'' \emph{CoRR}, vol.
  abs/2003.09514, 2020.

\bibitem{NEURIPS2019_9015}
A.~Paszke \emph{et~al.}, ``Pytorch: An imperative style, high-performance deep
  learning library,'' in \emph{Advances in Neural Information Processing
  Systems 32}.\hskip 1em plus 0.5em minus 0.4em\relax Curran Associates, Inc.,
  2019, pp. 8024--8035.

\bibitem{vishnevskiy2017isotropic}
V.~Vishnevskiy \emph{et~al.}, ``Isotropic total variation regularization of
  displacements in parametric image registration,'' \emph{IEEE transactions on
  medical imaging}, vol.~36, no.~2, pp. 385--395, 2017.

\bibitem{Schreiber_2022}
D.~Schreiber \emph{et~al.}, ``Crane: a 10 degree-of-freedom, tele-surgical
  system for dexterous manipulation within imaging bores,'' in \emph{2022
  International Conference on Robotics and Automation (ICRA)}, 2022, pp.
  5487--5494.

\bibitem{Jingbin_2021}
J.~Huang \emph{et~al.}, ``Model-predictive control of blood suction for
  surgical hemostasis using differentiable fluid simulations,'' in \emph{2021
  IEEE International Conference on Robotics and Automation (ICRA)}, 2021, pp.
  12\,380--12\,386.

\end{thebibliography}

\end{document}